\documentclass[aps, prd, floatfix,nofootinbib,superscriptaddress,amsmath,amssymb,amsfonts, notitlepage]{revtex4-1} 

\usepackage[T1]{fontenc}
\usepackage[latin9]{inputenc}
\usepackage{lmodern} 
\usepackage{color}
\usepackage{bbold}
\usepackage{dsfont}
\usepackage{graphicx}
\usepackage[caption=false]{subfig}
\usepackage[colorlinks]{hyperref}
\usepackage[bottom]{footmisc}
\usepackage{enumerate}
\usepackage{float}
\usepackage{mathtools}

\usepackage{empheq}
\usepackage{empheq}
\usepackage{tikz}

\makeatletter

\@ifundefined{textcolor}{}
{%
 \definecolor{BLACK}{gray}{0}
 \definecolor{WHITE}{gray}{1}
 \definecolor{RED}{rgb}{1,0,0}
 \definecolor{GREEN}{rgb}{0,1,0}
 \definecolor{BLUE}{rgb}{0,0,1}
 \definecolor{CYAN}{cmyk}{1,0,0,0}
 \definecolor{MAGENTA}{cmyk}{0,1,0,0}
 \definecolor{YELLOW}{cmyk}{0,0,1,0}
 }
\usepackage{fancyhdr}
\usepackage[normalem]{ulem}
\DeclareMathAlphabet\mbc{OMS}{cmsy}{b}{n}
\usepackage{balance}
\usepackage{MnSymbol}

\begin{document}

\global\long\def\eqn#1{\begin{align}#1\end{align}}
\global\long\def\vec#1{\overrightarrow{#1}}
\global\long\def\ket#1{\left|#1\right\rangle }
\global\long\def\bra#1{\left\langle #1\right|}
\global\long\def\bkt#1{\left(#1\right)}
\global\long\def\sbkt#1{\left[#1\right]}
\global\long\def\cbkt#1{\left\{#1\right\}}
\global\long\def\abs#1{\left\vert#1\right\vert}
\global\long\def\cev#1{\overleftarrow{#1}}
\global\long\def\der#1#2{\frac{{d}#1}{{d}#2}}
\global\long\def\pard#1#2{\frac{{\partial}#1}{{\partial}#2}}
\global\long\def\re{\mathrm{Re}}
\global\long\def\im{\mathrm{Im}}
\global\long\def\dd{\mathrm{d}}
\global\long\def\ddd{\mathcal{D}}
\global\long\def\seff{S_{\mathrm{M, IF}}}
\global\long\def\sint{S_{\mathrm{int}}}
\global\long\def\avg#1{\left\langle #1 \right\rangle}
\global\long\def\aavg#1{\left\llangle #1 \right\rrangle}
\global\long\def\mr#1{\mathrm{#1}}
\global\long\def\mb#1{{\mathbf #1}}
\global\long\def\mc#1{\mathcal{#1}}
\global\long\def\tr{\mathrm{Tr}}
\global\long\def\dbar#1{\Bar{\Bar{#1}}}

\global\long\def\nth{$n^{\mathrm{th}}$\,}
\global\long\def\mth{$m^{\mathrm{th}}$\,}
\global\long\def\non{\nonumber}

\newcommand{\orange}[1]{{\color{orange} {#1}}}
\newcommand{\cyan}[1]{{\color{cyan} {#1}}}
\newcommand{\blue}[1]{{\color{blue} {#1}}}
\newcommand{\yellow}[1]{{\color{yellow} {#1}}}
\newcommand{\green}[1]{{\color{green} {#1}}}
\newcommand{\red}[1]{{\color{red} {#1}}}
\global\long\def\todo#1{\orange{{$\bigstar$ \cyan{\bf\sc #1}}$\bigstar$} }

\title{Dissipative dynamics of a particle coupled to field via internal degrees of freedom}

\author{Kanupriya Sinha}
\email{kanu@princeton.edu}
\affiliation{Department of Electrical Engineering, Princeton University, Princeton, New Jersey 08544, USA}
\author{Adri\'{a}n Ezequiel Rubio L\'{o}pez}
\email{adrianrubiolopez0102@gmail.com}

\affiliation{Institute for Quantum Optics and Quantum Information of the Austrian Academy of Sciences,\\
Institute for Theoretical Physics,
University of Innsbruck, A-6020 Innsbruck, Austria}

\author{Yi\u{g}it Suba\c{s}\i}
\email{ysubasi@lanl.gov}
\affiliation{Computer, Computational and Statistical Sciences Division, Los Alamos National Laboratory, Los Alamos, NM 87545, USA}

\begin{abstract}

We study the non-equilibrium dissipative dynamics of the center of mass of a particle coupled to a field via its internal degrees of freedom. We model the internal and  external degrees of freedom of the particle as quantum harmonic oscillators in 1+1 D, with the internal oscillator coupled to a scalar quantum field at the center of mass position. Such a coupling results in a nonlinear interaction between the three pertinent degrees of freedom -- the center of mass, internal degree of freedom, and the field. It is typically assumed that the internal dynamics is  decoupled from that of the center of mass owing to their disparate characteristic time scales. Here we use an influence functional approach that allows one to account for the self-consistent backaction of the different degrees of freedom on each other, including the coupled non-equilibrium  dynamics of the internal degree of freedom and the field, and their influence on the dissipation and noise of the center of mass. Considering a weak nonlinear interaction term, we employ a perturbative generating functional approach to derive a second order effective action and a corresponding quantum Langevin equation describing the non-equilibrium dynamics of the center of mass. We analyze the resulting dissipation and noise  arising from  the field and the internal degree of freedom as a composite environment.  Furthermore, we establish a generalized fluctuation-dissipation relation for the late-time dissipation and noise kernels. Our results are pertinent to  open quantum systems that possess intermediary degrees of freedom between system and environment, such as in the case of optomechanical interactions.
\end{abstract}

\maketitle

\section{Introduction}


 Dissipative non-equilibrium dynamics of complex quantum systems  comprising different degrees of freedom is a subject of significant interest from the  perspective of theory of open quantum systems, as well as, emerging quantum information applications and devices   \cite{BPBook, Weiss, Clerk20}. While the dissipative dynamics of a reduced quantum system coupled linearly to its environment has been studied extensively, the case of a nonlinear coupling between a system and its environment via an intermediate degree of freedom and its effect on the resulting dynamics is seldom discussed \cite{HPZ93, Maghrebi16, Lampo16}. A commonly employed assumption  is that the time-scales associated with the different degrees of freedom  are sufficiently disparate so that their dynamics can be treated as being effectively decoupled from each other \cite{Stenholm86, Lan08, Haake}. Such an assumption precludes the possibility of studying the rich interplay between the different degrees of freedom, and its effect on the  non-equilibrium  dynamics of the system of interest. Moreover, an adiabatic elimination of fast variables does not allow one to see the effects of the non-equilibrium dynamics and quantum fluctuations of the fast degrees of freedom on the system dissipation and noise.

A canonical example of  such a complex open quantum system is the optomechanical interaction between a neutral polarizable particle  and a field, for instance, a moving atom,  nanoparticle, or  molecule interacting with the electromagnetic (EM) field \cite{Wieman, Yin, Molecules}. As an essential feature  these systems possess internal electronic  degrees of freedom that interact with the EM field and are located at the center of mass position represented by  the  mechanical degree of freedom (MDF).  While the MDF of the particle does not couple to the EM field directly, its internal degrees of freedom (IDF) interact with the EM  field at the center of mass position, thus   mediating an effective coupling between the MDF and the field. Considering the MDF as the system of interest, with the EM field and the IDF acting as an environment, such an effective interaction leads to  dissipation and noise in system dynamics. There is an interesting interplay between the three degrees of freedom  (the MDF, the IDF, and the EM field) that takes place in such a scenario as has been previously explored in \cite{KSthesis, MOF1, MOF2, Wang14}. Considering the dynamics from an approach that includes the self-consistent backaction of all the degrees of freedom on each other allows one to examine the role of IDF in the center of mass dynamics. For example, it has been shown that the IDF can facilitate a transfer of correlations between the field and center of mass in the case of optomechanical interactions \cite{KSthesis, MOF2} and affect the center of mass decoherence  \cite{Brun16}. On the other hand, it has also been demonstrated that including the quantized center of mass motion can affect the radiation reaction and thermalization of the IDF \cite{AERL19}.

We study here a minimal  model that captures the interplay between different degrees of freedom in a composite system and the nonequilibrium dissipative center of mass dynamics. Such a model forms the basis for optomechanical interactions between neutral particles and fields, and more generally applies to  open quantum systems that couple to an environment via internal degrees of freedom. The remainder of this paper is organized as follows. In Section \ref{model} we describe the model in consideration, in Section \ref{derivation} we describe the derivation of a second order  effective action  for the center of mass by integrating out the IDF and the field using a perturbative  influence functional approach \cite{HPZ93}. In Section \ref{Dynamics} we write an effective Langevin equation for the center of mass, discuss the resulting dynamics and present a generalized fluctuation-dissipation relation. We summarize our findings and present an outlook of the work in Section \ref{Discussion}.

\begin{figure}[t]
    \centering
    \includegraphics[width = 3.5 cm]{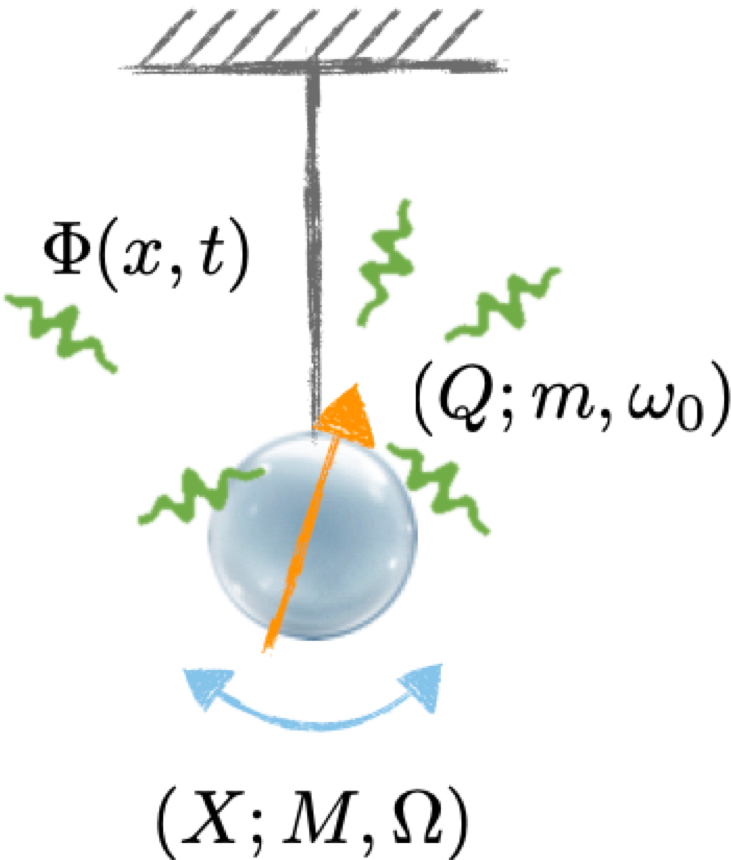}
    \caption{Schematic representation of a mechanical  oscillator of mass $M$ and frequency $\Omega $ coupled to a massless scalar field $\Phi$ via an  IDF. The internal oscillator with position coordinate $Q$ is described as a harmonic oscillator of mass $m $ and frequency $\omega_0 $ and couples to the scalar field $\Phi\bkt{{x},t} $ at the center of mass position ${X}$ of the mechanical oscillator.}
    \label{Schematic}
\end{figure}

\section{Model}
\label{model}


Let us consider a particle with its center of mass described by a mechanical oscillator of mass $M$ and frequency $\Omega$, and its polarization described by an IDF as a harmonic oscillator of mass $m$ and resonance frequency $\omega_0$. The composite system is assumed to be interacting with a quantum field, which we consider to be a scalar field $\Phi(x,t)$ in  $1+1$ D. The total action of the  system can be written as
\eqn{
S = S_{M}+S_{I}+S_{F}+S_{\mathrm{int},}
\label{TotalAction}
}
where 
\eqn{S_M &= \int\dd\tau\bkt{ \frac{1}{2}M \dot{X}^2-\frac{1}{2}M\Omega^2 X^2}}
refers to the free action for the mechanical oscillator  with $\cbkt{X, \dot {X}}$ referring to the center of mass position and velocity, while
\eqn{
S_I &= \int\dd\tau\bkt{ \frac{1}{2}m \dot{Q}^2-\frac{1}{2}m\omega_0^2 Q^2}}
corresponds to the free action for the internal degree of freedom with amplitude $Q$, and
\eqn{
S_F &= \int\dd\tau\int\dd{x} \frac{\epsilon_0}{2}\sbkt{\bkt{\partial_t\Phi(x,\tau)}^2- \bkt{\mb{\partial_x}\Phi(x,\tau)}^2}}
is the free action for the scalar field $\Phi\bkt{x,t}$, $\epsilon_0 $  being the vacuum permittivity. We  consider $ \hbar = c = 1$ throughout this paper. The correspondence with the EM field is established by identifying the scalar field $\Phi\bkt{{x},t}$ as the vector potential. 

 Considering that the particle interacts with the field via its IDF, at the position determined by the MDF, the interaction action is given as
\eqn{
S_{\mathrm{int}} &= \int\dd\tau \int \dd{x} \lambda  Q\Phi({x},\tau)\delta\bkt{{x}-X(\tau)}
}
with the strength of the interaction defined by $\lambda$. We note that the above interaction action is nonlinear, with the two degrees of freedom and the field interacting together. 
Such a model, previously referred to as the mirror-oscillator-field model has been studied in the context of describing optomechanical interactions from a microscopic perspective \cite{MOF1, MOF2, KSthesis}. It provides a self-consistent treatment of the different degrees of freedom involved, in that one does not need to impose the boundary conditions on the field by hand,  and the mechanical effects of the field are consistently described upon tracing over the IDF.

In typical experimental systems of relevance, such as those of atoms and nanoparticles confined in traps,  the center of mass motion is restricted to a region of the space smaller than the wavelengths of the field that interact resonantly with the internal degrees. Thus motivated by practical considerations, we assume that the  motion of the MDF is restricted to small displacements about the average equilibrium position  $X_0$, such that we can expand the interaction action to first order in displacements from the equilibrium as follows 
\eqn{ S_\mr{int}&\approx\int\dd\tau  \sbkt{\lambda  Q\Phi(X_0,\tau)+\lambda  Q\partial_x\Phi(X_0,\tau) \bkt{X - X_0}}.
\label{IntAction}}
The first term in the above action represents the linear interaction between the IDF and the field, similar to the $\sim \mb{p}\cdot\mb{A}$ coupling in atom-field interactions. The second term stands for the tripartite interaction between the IDF, the MDF and the field. Thus, as we show in the following, the effective coupling of the center of mass to the field via the IDF leads to its dissipative dynamics. From now on, for simplicity and without loss of generality, we assume $X_0=0$.

\subsection{Scalar field as bath of harmonic oscillators}

The above separation of the total interaction action (Eq.\eqref{IntAction}) suggests splitting the field into two separate oscillator baths such that one of the baths interacts with the IDF, while the other is coupled to both the internal and mechanical degrees.  We thus make an eigenmode decomposition of the field to rewrite it in terms of a bath of oscillators as follows \cite{QBM3}
\eqn{
\Phi({x},\tau) = \sum_n\sqrt{ \frac{1}{\epsilon_0{L}}} \sbkt{q_n^{(-)} \cos({\omega}_n{x})+q_n^{(+)} \sin({\omega}_n{x})},
}
where $L$ is the mode volume of the field, $\omega_n$ refers to the mode frequency, and $q_n^{(+)}$ and $q_n^{(-)}$ refer to two independent set of eigenmodes of the field, assuming periodic boundary conditions.  One can thus rewrite the free field action as as a sum $S_F = S_F ^{(+)}+S_F ^{(-)} $, where 
\eqn{
S_F^{(\pm)} = \int \dd \tau\sum_n \frac{1}{2} \sbkt{\left.\dot{q}_n^{(\pm)}\right.^2-\omega_n^2\left.{q}_n^{(\pm)}\right.^2}
}
with $S_F^{(\pm)}$ corresponding to two separate baths of $(+)$ and $(-)$ oscillators.
Thus, for $X_0 =0$ the interaction action in Eq.\eqref{IntAction} can be written as a sum of two separate interaction terms $ S_{\mathrm{int}} = S_{\mathrm{int}}^{(-)}+S_{\mathrm{int}}^{(+)} $, with 
\eqn{
\label{Sint-}
S_{\mathrm{int}} ^{(-)}&=  \int\dd\tau\lambda Q\sum_nq_n^{(-)}\\
\label{Sint+}
S_{\mr{int}}^{(+)}&=\int\dd\tau\lambda Q\sum_n\omega_nX q_n^{(+)} .
}
Thus the interaction action $S_\mr{int}^{(-)}$  linearly couples the $(-)$ bath of oscillators of the scalar field to the IDF,  leading to the dissipative dynamics of  the IDF. The interaction action $S_\mr{int}^{(+)}$ couples the center of mass to the $(+)$ bath of oscillators  via the IDF. We note that $S_\mr{int}^{(+)}$ is essentially responsible for all the mechanical effects of the field, such as radiation pressure due to optical fields and vacuum-induced forces \cite{PierreBook, PeterQOBook}. For instance,  one can see that eliminating the IDF  from the interaction action $S_\mr{int}^{(+)}$ yields an effective nonlinear interaction between the MDF and the field, similar to the intensity-position coupling in optomechanical interactions \cite{KSthesis}. We also remark here that the above separation of the field into two uncorrelated baths is akin to the Einstein-Hopf theorem that establishes the statistical independence of the blackbody radiation field and its derivative \cite{EinHopf1, EinHopf3}.

\section{Effective action derivation from perturbative influence functional approach}
\label{derivation}

The nonlinearity of the interaction action $S_\mr{int} ^{(+)}$  limits the possibility of obtaining an exact analytical solution for the center of mass dynamics. We therefore follow a perturbative generating functional approach as prescribed in \cite{HPZ93}, assuming that the characteristic nonlinear coupling strength $\bkt{\sim \lambda \omega_n}$ can be treated perturbatively to write the resulting dynamics of the center of mass.

Let us consider the evolution of the system density matrix $\hat \rho_M $ as follows 
\eqn{\label{rhomt}&\hat \rho_M\bkt{t}= \tr_I \tr_{F^+}\tr_{F^-}\sbkt{\hat{\mc{U}}(t,0)\bkt{\hat \rho_M(0)\otimes\hat \rho_I(0)\otimes\hat \rho_F^{(+)}(0)\otimes \hat\rho_F^{(-)}(0)}\hat{\mc{U}}^\dagger(t,0)},}
where we have assumed that the density operators for the different degrees of freedom are initially uncorrelated with each other, with $\hat \rho_I $ and $\hat \rho_F^{(\pm)}$ referring to the density matrices for the IDF and the $(\pm)$ bath, and $ \hat{\mc{U}}(t,0)$ corresponding to the time evolution operator. We write the density matrices in a coordinate representation such that

\eqn{\hat{\rho}_M (\tau) = &\int \dd X \dd X' \rho_M\bkt{X,X';\tau} \ket{X}\bra{X'},\\
\hat{\rho}_I (\tau) = &\int \dd Q \dd Q' \rho_I\bkt{Q,Q';\tau} \ket{Q}\bra{Q'},\\
\hat{\rho}_{F}^{(\pm)} (\tau) = &\prod_n \int \dd q_n ^{(\pm)} \dd {q^{(\pm)}_n}' \rho_F^{(\pm)}\bkt{\cbkt{ q_n ^{(\pm)}, {q^{(\pm)}_n}'};\tau} \ket{ q_n^{(\pm)} }\bra{ {q^{(\pm)}_n}'},} such that the system evolution in Eq.\,\eqref{rhomt} can be rewritten as
%
\begin{widetext}
\eqn{
\rho_M\bkt{{X}_f,{X}_f';t}=&\prod_{m,n} \int\dd Q_f \dd q^{(+)}_{nf} \dd q^{(-)}_{mf}\int\dd {X}_i\dd Q_i \dd q^{(+)}_{ni} \dd q^{(-)}_{mi}\int\dd {X}_i'\dd Q_i' \dd {q^{(+)}_{ni}}' \dd {q^{(-)}_{mi}}'
\non\\
&\sbkt{\rho_M\bkt{{X}_i,{X}_i';0}\otimes\rho_I\bkt{Q_i,Q_i';0}\otimes\rho_F\bkt{\cbkt{q^{(+)}_{ni},{q^{(+)}_{ni}}'};0}\otimes\rho_F\bkt{\cbkt{q^{(-)}_{mi},{q^{(-)}_{mi}}'};0}\right.\non\\
&\left.\mc{J}\bkt{X_f, Q_f , \cbkt{ {q_{nf}^{(+)}}, {q_{mf}^{(-)}}};t\big\vert X_i, Q_i , \cbkt{ {q_{ni}^{(+)}}, {q_{mi}^{(-)}}};0}\mc{J}^\dagger\bkt{X'_f, Q_f ,\cbkt{ {q_{nf}^{(+)}}, {q_{mf}^{(-)}}};t\big\vert X'_i, Q'_i ,\cbkt{ {q_{ni}^{(+)}}', {q_{mi}^{(-)}}'};0}},
}
\end{widetext}
where  $\cbkt{{X}_i, {X}_i'}$ and $\cbkt{ {X}_f, {X}_f'}$ refer to the initial and final coordinates corresponding to the center of mass variables   respectively, $\cbkt{ Q_i,Q_i'; Q_f,  Q_f' }$ are those for the internal oscillator, and $\cbkt{ q_{ni}^{(\pm)} , {q_{ni}^{(\pm)}}' ; q_{nf}^{(\pm)}, {q_{nf}^{(\pm)}}'}$ are the initial and final coordinates for the $n^\mr{th}$ oscillator of the $(\pm)$ bath. We  assume that the initial states  $\rho_F^{(\pm)}\bkt{\cbkt{q^{(\pm)}_{ni},{q^{(\pm)}_{ni}}'};0}$ of the field and $\rho_I\bkt{Q_i,Q_i';0}$ of the IDF are  thermal with a temperature $T_F$ and $T_I$ respectively.  The forward propagator is defined as $\mc{J}\bkt{X_f, Q_f , \cbkt{ q_{nf}^{(+)}, q_{mf}^{(-)}};t\big\vert X_i, Q_i ,\cbkt{ q_{ni}^{(+)}, q_{mi}^{(-)}};0}\equiv \prod_{m,n} \bra{{X}_f}\bra{Q_f}\bra{q^{(+)}_{nf}}\bra{q^{(-)}_{mf}}\hat{\mc{U}} \bkt{t,0}\ket{q^{(-)}_{mi}}\ket{q^{(+)}_{ni}}\ket{Q_i}\ket{{X}_i}$.  Thus the first and last terms in the integral refer to the forward and backward propagators that can be expressed in  path integral representation to write the evolution as
\eqn{
&\rho_M\bkt{{X}_f,{X}_f';t}=\int\dd {X}_i\int\dd {X}_i'\,\rho_M\bkt{{X}_i,{X}_i';0}\int_{{X}(0) = {X}_i}^{{X}(t) = {X}_f}\ddd {X}\int_{{X}'(0) = {X}'_i}^{{X}'(t) = X'_f}\ddd {X}' e^{i\bkt{S_M[{X}] - S_M[{X}']}}\mc{F}\sbkt{{X},{X}'},
\label{rhoMx}
}
where $\mc{F}\sbkt{{X}, {X}'}$ refers to the  influence functional  of the field and the internal oscillator  acting on the MDF \cite{FeynmanTrick}, which can be written explicitly as
\eqn{\label{IFxx'}
&\mc{F}\sbkt{{X},{X}'}\equiv\non\\
&\prod_{m,n} \int\dd Q_f\dd q^{(+)}_{nf} \dd q^{(-)}_{mf} \int \dd Q_i \dd q^{(+)}_{ni} \dd q^{(-)}_{mi}\int \dd Q_i' \dd {q^{(+)}_{ni}}'\dd {q^{(-)}_{mi}}'\rho_I\bkt{Q_i,Q_i';0}\otimes\rho_F^{(+)}\bkt{\cbkt{q^{(+)}_{ni},{q^{(+)}_{ni}}'};0}\otimes \rho_F^{(-)}\bkt{\cbkt{q^{(-)}_{mi},{q^{(-)}_{mi}}'};0}\non\\
&\int_{Q(0) = Q_i}^{Q(t) = Q_f}\ddd Q\int_{Q'(0) = Q'_i}^{Q'(t) = Q_f}\ddd Q' e^{i\bkt{S_I[Q] - S_I[Q']}}
\int_{q^{(+)}_n(0) = q^{(+)}_{ni}}^{q^{(+)}_{n}(t) = q^{(+)}_{nf}}\ddd q_n^{(+)}\int_{{q^{(+)}_n}'(0) = {q^{(+)}_{ni}}'}^{{q^{(+)}_{n}}'(t) = q^{(+)}_{nf}}\ddd {q^{(+)}_n}' e^{i\bkt{S_{F}^{(+)}\sbkt{\cbkt{q_n^{(+)}}} - S_{F}^{(+)}\sbkt{\cbkt{{q_n^{(+)}}'}}}} \non\\
&\int_{q^{(-)}_m(0) = q^{(-)}_{mi}}^{q^{(-)}_{m}(t) = q^{(-)}_{mf}}\ddd q_m^{(-)}\int_{{q^{(-)}_m}'(0) = {q^{(-)}_{mi}}'}^{{q^{(-)}_{m}}'(t) = q^{(-)}_{mf}}\ddd {q^{(-)}_m}' e^{i\bkt{S_{F}^{(-)}\sbkt{\cbkt{q_m^{(-)}}} - S_{F}^{(-)}\sbkt{\cbkt{{q_m^{(-)}}'}}}}e^{i\bkt{S_{\mr{int}}^{(-)}\sbkt{  Q,\cbkt{q_m^{(-)}}} - S_{\mr{int}}^{(-)}\sbkt{  Q',\cbkt{{q_m^{(-)}}'}}} }\non\\
&e^{i\bkt{S_{\mr{int}}^{(+)}\sbkt{{X},  Q,\cbkt{q_n^{(+)}}} - S_{\mr{int}}^{(+)}\sbkt{{X}',  Q',\cbkt{{q_n^{(+)}}'}}}}.
}
%

 Thus we have treated here the MDF as the system and the IDF and the field as the bath, with the influence functional capturing the influence of the bath on the evolution of the MDF. We will now evaluate the influence functional in a perturbative manner, by first tracing out the $(-)$ bath modes that couple only to the IDF, and then the $(+) $ bath modes and the IDF that couple to the center of mass.

\subsection {Tracing over the $(-)$ bath}


Let us define ${\mc{F}^{(-)}\sbkt{Q,  Q'}}$ as the influence functional that accounts for the influence of the $(-)$ bath on the IDF as follows
\eqn{\label{IF-}
\mc{F}^{(-)}\sbkt{  Q,  Q'} \equiv \prod_m\int\dd q^{(-)}_{mf}\int\dd q^{(-)}_{mi}\dd {q^{(-)}_{mi}}'\rho_F^{(-)}\bkt{\cbkt{q^{(-)}_{mi},{q^{(-)}_{mi}}'};0}\int_{q^{(-)}_m(0) = q^{(-)}_{mi}}^{q^{(-)}_{m}(t) = q^{(-)}_{mf}}\ddd q_m^{(-)}\int_{{q^{(-)}_m}'(0) = {q^{(-)}_{mi}}'}^{{q^{(-)}_{m}}'(t) = {q^{(-)}_{mf}}'}\ddd {q^{(-)}_m}'\non\\e^{i\bkt{S_{F}^{(-)}\sbkt{\cbkt{q_m^{(-)}}} - S_{F}^{(-)}\sbkt{\cbkt{{q_m^{(-)}}'}}}}e^{i\bkt{S_{\mr{int}}^{(-)}\sbkt{  Q,\cbkt{q_m^{(-)}}} - S_{\mr{int}}^{(-)}\sbkt{  Q',\cbkt{{q_m^{(-)}}'}}}}.
}
Performing the path integrals by considering an initial thermal state of the $(-) $ bath oscillators with temperature $T_F$ corresponding to the temperature of the field, one obtains \cite{CalzettaHu}
\eqn{\label{Fexp-}
\mc{F}^{(-)}\sbkt{  Q,  Q'}=  e^{{i}S^{(-)}_{\mr{I, IF}}\sbkt{  Q,  Q'}},
}
where the influence action due to the $(-)$ bath is given as
\eqn{
S^{(-)}_{\mr{I, IF}}\sbkt{  Q, Q'}\equiv-\int_0^t\dd \tau_1\int_0^{\tau_1}\dd\tau_2 \sbkt{  Q\bkt{\tau_1} -  Q'\bkt{\tau_1} }\eta^{(-)}\bkt{\tau_1 - \tau_2} \sbkt{  Q \bkt{\tau_2} +  Q' \bkt{\tau_2}}\non\\
+i\int_0^t\dd \tau_1\int_0^{\tau_1}\dd\tau_2 \sbkt{  Q\bkt{\tau_1} -  Q'\bkt{\tau_1} }\nu^{(-)}\bkt{\tau_1 - \tau_2} \sbkt{  Q \bkt{\tau_2} -  Q' \bkt{\tau_2}}.
\label{smineff}
}
The above influence action corresponds to the case of quantum Brownian motion (QBM) model with a linear system-bath coupling \cite{HPZ92}. The dissipation and noise kernels, $\eta^{(-)}(\tau)$ and $\nu^{(-)}(\tau)$  respectively, are
\eqn{
\eta^{(-)}(\tau) =&- \sum_n \frac{\lambda^2 }{2 \omega_n }\sin\bkt{\omega_n \tau}{\Theta(\tau)}\label{DissQBMMinus}\\
\nu^{(-)}(\tau) =& \sum_n \frac{\lambda^2}{2 \omega_n }\coth\bkt{\frac{ \omega_{n}}{2k_BT_F}}\cos\bkt{\omega_n \tau}.\label{NoiseQBMMinus}
}
The spectral density associated with the $(-)$ bath is thus
\eqn{\label{J-}
J^{(-)}\bkt{\omega} = \sum_n \frac{\lambda^2 }{2 \omega_n } \delta\bkt{\omega - \omega_n }.
} We note from the above that the influence of the $(-)$ bath on the IDF leads to its QBM dynamics, with a sub-Ohmic spectral density, which necessitates a non-Markovian treatmeant of the resulting dynamics \cite{Weiss}.  

We now consider the effect of the IDF in turn on the dynamics of the center of mass. Using Eq.\,\eqref{IF-} and \eqref{Fexp-}, we can rewrite the total influence functional in Eq.\,\eqref{IFxx'} as
\eqn{\label{Fxx'}
\mc{F}\sbkt{{X},{X}'}=\prod_{n} &\int\dd Q_f\dd q^{(+)}_{nf}\int \dd Q_i\dd q^{(+)}_{ni} \int \dd Q_i' \dd {q^{(+)}_{ni}}'\,\rho_I\bkt{Q_i,Q_i';0}\otimes\rho_F^{(+)}\bkt{\cbkt{q^{(+)}_{ni},{q^{(+)}_{ni}}'};0}\non\\
&\int_{Q(0) = Q_i}^{Q(t) = Q_f}\ddd Q\int_{Q'(0) = Q'_i}^{Q'(t) = Q'_f}\ddd Q' e^{i\bkt{S_I[Q] - S_I[Q']}}
\int_{q_n^{(+)}(0) = q_{ni}^{(+)}}^{q_{n}^{(+)}(t) = q_{nf}^{(+)}}\ddd q_n\int_{{q^{(+)}_n}'(0) = {q^{(+)}_{ni}}'}^{{q^{(+)}_n}'(t) = q^{(+)}_{nf}}\ddd {q^{(+)}_n}'  \non\\
&e^{i\bkt{S_{F}\sbkt{\cbkt{q_n^{(+)}}} - S_{F}\sbkt{\cbkt{{q_n^{(+)}}'}}}} e^{i\bkt{S_{\mr{int}}^{(+)}\sbkt{{X},Q,\cbkt{q_n^{(+)}}} - S^{(+)}_{\mr{int}}\sbkt{{X}',  Q',\cbkt{{q_n^{(+)}}'}}}}e^{i S_{\mr{I, IF}}^{(-)}\sbkt{Q,Q'}},
}
where the integrations over the IDF and the $(+)$ bath of oscillators remain. Thus far the above influence functional is exact in that we have not made any additional approximations with regard to the strength of coupling when tracing over the $(-)$ bath.

\subsection{Tracing over the $(+)$ bath and the internal degree of freedom}\label{TracingInternalDOF}

Next we would like to integrate out the (+) bath and the IDF, which are nonlinearly coupled to the system, using a perturbative generating functional approach.  The generating functional is simply the influence functional of the environment where the bath oscillators are linearly coupled to the system.  We define the influence action $S_\mr{M,IF}\sbkt{X,X'}$ corresponding to the influence functional $\mc F\sbkt{{X},{X}'}$ in Eq.\eqref{Fxx'} such that $\mc F\sbkt{{X},{X}'}\equiv e^{i S_\mr{M,IF}}$. The influence action up to second order in the  coupling strength $\lambda$ can be obtained  as (see appendix \ref{App:seff} for a detailed derivation) \cite{HPZ93}:
\eqn{\label{seff}
\seff^{(2)}[{X},{X}'] \approx& \avg{\sint^{(+)}\sbkt{{X},\frac{1}{i}\frac{\delta}{\delta {\bf J}}}}_0 - \avg{\sint^{(+)}\sbkt{{X}',-\frac{1}{i}\frac{\delta}{\delta {\bf J'}}}}_0\non\\
&+\frac{i}{2}\bkt{ \avg{\cbkt{\sint^{(+)}\sbkt{{X},\frac{1}{i}\frac{\delta}{\delta {\bf J}}}}^2}_0 - \cbkt{\avg{\sint^{(+)}\sbkt{{X},\frac{1}{i}\frac{\delta}{\delta {\bf J}}}}_0}^2 }\nonumber\\
& +\frac{i}{2}\bkt{ \avg{\cbkt{\sint^{(+)}\sbkt{{X}',-\frac{1}{i}\frac{\delta}{\delta {\bf J'}}}}^2}_0 - \cbkt{\avg{\sint^{(+)}\sbkt{{X}',-\frac{1}{i}\frac{\delta}{\delta {\bf J'}}}}_0}^2}\nonumber\\
 &-i\bkt{ \avg{\sint^{(+)}\sbkt{{X},\frac{1}{i}\frac{\delta}{\delta {\bf J}}}\sint^{(+)}\sbkt{{X}',-\frac{1}{i}\frac{\delta}{\delta {\bf J'}}}}_0 - \avg{\sint^{(+)}\sbkt{{X},\frac{1}{i}\frac{\delta}{\delta {\bf J}}}}_0\avg{\sint^{(+)}\sbkt{{X}',-\frac{1}{i}\frac{\delta}{\delta {\bf J'}}}}_0 },}
 where we have defined $ {\bf J}\equiv \bkt{J,\cbkt{J_n}}$, $ {\bf J'}\equiv \bkt{J',\cbkt{J'_n}}$, and the expectation values $\avg{\mc{O}[{J},{J'}]}_0\equiv\left. \mc{O}[{J},{J'}]\mc G^{(1)}[{J},{J'}]\right\vert_{{ J} = { J'}=0}$. The  generating functional  $\mc G^{(1)}[{J},{J'}]$  for the $(+) $ bath and the IDF is defined as
\eqn{\label{Gentot}
\mc{G}^{(1)}[J,J',\cbkt{J_n,J'_n}] \equiv\prod_n  \mc{F}^{(1)}_n\sbkt{J_n,J'_n} \mc{F}^{(1)}_I\sbkt{J,J'}
}
with $\mc{F}^{(1)}_I\sbkt{J,J'}$ and $\mc{F}^{(1)}_n\sbkt{J_n,J'_n}$ as the influence functionals for the internal oscillator and the $n^\mr{th}$ oscillator of the $(+)$ bath respectively, with a linear coupling to the  corresponding source current terms $\cbkt{J,J'}$ and $\cbkt{J_n,J'_n}$. An explicit derivation of the generating functional is given in Appendix\,\ref{App:gf}.

We can calculate the second order influence action in Eq.\,\eqref{seff} as shown in Appendix\,\ref{Seffavg} to obtain a form similar to that of linear QBM dynamics \cite{HPZ92}:
\begin{widetext}
\eqn{\seff^{(2)}\sbkt{X,X'} =& -\int_0^t\dd t_1\int_0^{t_1}\dd t_2 \,\sbkt{X(t_1)-X'(t_1)} \eta^{(2)} (t_1,t_2)\sbkt{X(t_2)+X'(t_2)}\nonumber\\
&+i\int_0^t\dd t_1\int_0^{t_1}\dd t_2 \, \sbkt{X(t_1)-X'(t_1)} \nu ^{(2)}(t_1,t_2)\sbkt{X(t_2)-X'(t_2)} ,
\label{EffAct2nd}}
\end{widetext}
where dissipation and noise kernels are defined as
\eqn{
\eta^{(2)} (t_1,t_2) =&\frac{1}{2}\eta ^{(+)} \bkt{t_1 - t_2} \cbkt{\nu_{GG} \bkt{ t_1,t_2} + \aavg{Q_h\bkt{ t_1} Q_h\bkt{t_2}}}+ 
\frac{1}{4}\nu ^{(+)} \bkt{t_1 - t_2}{g}\bkt{t_1 - t_2} \Theta\bkt{t_1 - t_2},\label{Eta2}\\
\nu^{(2)} (t_1,t_2) =&  \frac{1}{2}\nu ^{(+)} \bkt{t_1 - t_2} \cbkt{\nu_{GG} \bkt{ t_1,t_2} + \aavg{Q_h\bkt{ t_1} Q_h\bkt{t_2}}}-\frac{1}{4}
\eta ^{(+)} \bkt{t_1 - t_2}{g}\bkt{t_1 - t_2} ,\label{Nu2}
}
{with ${G}(t)=g(t)\Theta(t)$} the propagator for the IDF defined as in Eq.\eqref{Gret}. The noise correlation of the IDF is given by \eqn{&\nu_{GG}(t_1,t_2)\equiv \int_0^t\dd\tau_1\int_0^t\dd\tau_2 {G}\bkt{t_1-\tau_1}\nu^{(-)}\bkt{\tau_1-\tau_2}{G}\bkt{t_2-\tau_2}.
\label{NuGG}
}
The noise arising from the dispersion in the initial
conditions of the IDF is captured  in the term $\sim \aavg{Q_h \bkt{t_1}Q_h \bkt{t_2}}$, where $Q_h\bkt{t}$ is the classical solution to the homogeneous Langevin equation (Eq.\,\eqref{IDFhom}) corresponding to the dissipative dyanmics of IDF in the presence of the $ (-)$ bath. The average $\aavg{\dots}$ is taken over  initial position and momentum distribution of the IDF, as defined in Eq.\,\eqref{aavg}.

The dissipation and noise kernels associated with the $(+)$ bath, defined based on a bilinear system-bath coupling are 
\eqn{\label{ETA+}
\eta^{(+)}\bkt{\tau} \equiv& - \frac{1}{2}\sum_n \lambda^2 \omega_n  \sin\bkt{\omega_n \tau}{\Theta(\tau)}\\
\label{NU+}
\nu^{(+)}\bkt{\tau}\equiv &  \frac{1}{2 }\sum_n \lambda^2 \omega_n  \coth\bkt{\frac{ \omega_{n}}{2k_BT_F}}\cos\bkt{\omega_n \tau},
}
 similar to those of the $(-)$ bath (Eq.\,\eqref{DissQBMMinus} and Eq.\,\eqref{NoiseQBMMinus}) with a coupling $\lambda\rightarrow \lambda \omega_n $. The spectral density associated with the $(+) $ bath is thus 
\eqn{\label{J+}
J^{(+)}\bkt{\omega} = \sum_n \frac{\lambda^2 \omega_n}{2}\delta \bkt{\omega - \omega_n},
}
corresponding to an Ohmic case \cite{CalzettaHu}. We note that while the two baths $(+)$ and $(-)$  correspond to the same field, due to a gradient coupling of the center of mass to the field the $(+)$ bath has an Ohmic spectral density while the $(-)$ bath is sub-Ohmic (see Eq.\,\eqref{J-}).

One can note a few important features from the influence action (Eq.\,\eqref{EffAct2nd}) and the dissipation and noise kernels therein (Eq.\,\eqref{Eta2} and Eq.\,\eqref{Nu2}):

\begin{itemize}
    \item {The overall structural similarity of the second order influence action to that of the linear QBM extends to structure of the dissipation and noise kernels as well. Meaning that writing the interaction Hamiltonian as $H \sim \hat X \hat {\mc{B}} $, where $\hat {\mc{B}} $ is the total bath operator (possibly nonlinear), one can write an effective action of the same form as linear QBM with dissipation and noise kernels given in terms of the expectation values of the  commutator and anticommutator of the two-time bath correlation functions $ \eta\bkt{\tau} = \avg{\sbkt{\hat{ \mc{B} }\bkt{\tau }, \hat {\mc{B}} \bkt{0}}}$ and $\nu\bkt{\tau} = \avg{\cbkt{\hat {\mc{B}} \bkt{\tau }, \hat {\mc{B}} \bkt{0}}}$. This is a more general property of bilinear system-bath couplings. The influence functional of Eq.\eqref{Fxx'} for $X=0=X'$ results in a Gaussian path integral, so it is reasonable to obtain that the second order results with a linear QBM form.  One can further verify as a check that the unitarity of the evolution implies  $\seff^{(2)}\sbkt{X,X}=0$}.
    \item{The dissipation kernel contains two parts that can be interpreted as: (1) coming from the dissipation kernel of the $(+)$ bath  $(\eta^{(+)}(t_1-t_2) )$ combined with the noise from the IDF $(\nu_{GG} \bkt{t_1, t_2}+ \aavg{Q_h\bkt{ t_1} Q_h\bkt{t_2}}$, and (2) linear propagator of the IDF $( G \bkt{t_1 - t_2})$ combined with the noise from the $(+)$ bath $(\nu^{(+)} \bkt{t_1 - t_2})$. Notice that the quantity $\nu_{GG} \bkt{ t_1,t_2} + \aavg{Q_h\bkt{ t_1} Q_h\bkt{t_2}}$ stands for the full quantum correlations of the IDF under the influence of the $(-)$ bath alone. The term $\aavg{Q_h\bkt{ t_1} Q_h\bkt{t_2}}$ accounts for the relaxation of the initial conditions' contribution, thus carrying the information about the initial state of the IDF and determined only by the dissipation caused by the $(-)$ bath. On the other hand, the term $\nu_{GG} \bkt{ t_1,t_2}$ stands for the fluctuations on the IDF generated by the fluctuations of the $(-)$ bath.}
    
    \item{The noise kernel has the combined noise from both the IDF $(\nu_{GG} \bkt{t_1, t_2}+ \aavg{Q_h\bkt{ t_1} Q_h\bkt{t_2}}$ and that from the $(+)$ bath $(\nu^{(+)} \bkt{t_1 - t_2})$. In addition to that the noise term also contains the two dissipation kernels from the IDF and the $(+)$ bath. }
    
\end{itemize}

\section {Langevin equation and fluctuation-dissipation relation}
\label{Dynamics}

Having determined the influence action for the center of mass, we can now turn to  studying its dynamics. This section is devoted to the deduction of an effective Langevin equation of motion for the center of mass and a corresponding generalized fluctuation-dissipation relation  in the long-time limit.

\subsection{Effective equation of motion for the center of mass}


Considering the influence action  obtained as a result of tracing out the IDF and the field (Eq.\eqref{EffAct2nd}), the total effective action for the center of mass variables up to second order reads:
\eqn{
S_{\mathrm{M, eff}}\sbkt{X,X'} = S_{M}\sbkt{X} - S_{M}\sbkt{X'} + \seff^{(2)}\sbkt{X,X'}.}
The above action is quadratic and has the same form as in the case of linear QBM \cite{CalzettaHu}. One can thus deduce an equation of motion for the center of mass by extremizing the above action  to obtain
\eqn{
\ddot{X}\bkt{t} + \Omega^{2}X\bkt{t} + \frac{2}{M} \int_{0}^{t}d\tau ~\eta^{(2)}(t,\tau)~X(\tau)=0.
\label{EffEqCOMAvg}}

A Langevin equation can be worked out  by implementing the Feynman and Vernon procedure for Gaussian path integrals \cite{FeynmanTrick,Behunin1}.  First, we notice that $i \seff^{(2)}=i {\rm Re}[\seff^{(2)}]-{\rm Im}[ \seff^{(2)}]$, and that both, the real and imaginary parts of the effective action are quadratic functionals of $\{X,X'\}$. Thus, ${\rm exp}(-{\rm Im}[\seff^{(2)}])$ is a Gaussian functional since the kernel owing to the quadratic nature of the action. Considering that a Gaussian functional can be written in terms of its functional Fourier transform (which is also a Gaussian functional), we can write ${\rm exp}(-{\rm Im}[\seff^{(2)}])$ as an influence functional over a new variable $\xi(t)$
\eqn{
e^{i\seff^{(2)}\sbkt{X,X'}}=\int\mathcal{D}\xi~\mathcal{P}\sbkt{\xi}
e^{i\int_{0}^{t}dt_{1}\sbkt{X(t_{1})-X'(t_{1})}\sbkt{\xi(t_{1})-\int_{0}^{t_{1}}dt_{2}\eta^{(2)}(t_{1},t_{2})\sbkt{X(t_{2})+X'(t_{2})}}},
\label{Seff2StochasticInt}
}
where the distribution of the new functional variable is given by
\eqn{
\mathcal{P}\sbkt{\xi}=e^{-\int_{0}^{t}dt_{1}\int_{0}^{t_{1}}dt_{2}\xi(t_{1})\sbkt{4\nu^{(2)}(t_{1},t_{2})}^{-1}\xi(t_{2})}.}

It is worth noting that the new variable $\xi\bkt{t}$ comes to replace the kernel describing the quantum and thermal fluctuations of the composite environment and drive the dynamics as an external stochastic force. To recover the influence action as in Eq.\,\eqref{Seff2StochasticInt}, one needs to integrate over $\xi$ given the functional distribution $\mathcal{P}[\xi]$, which is positive definite since the noise kernel $\nu^{(2)}\bkt{t_1 , t_2}$ is symmetric and positive. Thus, the stochastic variable $\xi\bkt{t}$ acts as a classical fluctuating force, which can be interpreted as a noise with a probability distribution  $\mathcal{P}[\xi]$. Furthermore, due to the  Gaussianity of $\mathcal{P}\sbkt{\xi\bkt{t}}$ the noise is completely characterized by its first and second moments:
\eqn{\label{ximom}&\avg{\xi\bkt{t}}_{\xi}=0, \non\\
&\avg{\xi(t_{1})\xi(t_{2})}_{\xi}=4\nu^{(2)}(t_{1},t_{2}),}  where $\avg{...}_{\xi}=\int\mathcal{D}\xi\mathcal{P}\sbkt{\xi}(...)$. Now, we can re-define an effective action for the center of mass including this new variable as:
\eqn{
\tilde{S}_{\mathrm{M, eff}}\sbkt{X,X',\xi} = S_{M}\sbkt{X} - S_{M}\sbkt{X'} + {\mathrm{Re}}\sbkt{\seff^{(2)}\sbkt{X,X'}}
+\int_{0}^{t}dt_{1}\sbkt{X(t_{1})-X'(t_{1})}\xi(t_{1}).}

Finally, associated with this new action we can derive an effective equation of motion for the center of mass $(\delta \tilde{S}_{\mathrm{M, eff}}/\delta X_{\Delta})|_{X_{\Delta}=0}=0$, which now gives the Langevin equation:
\eqn{
\ddot{X}\bkt{t} + \Omega^{2}X\bkt{t} + \frac{2}{M} \int_{0}^{t}d\tau ~\eta^{(2)}(t,\tau)~X(\tau)=\xi\bkt{t},
\label{EffEqCOMLangevin}}
subjected to the statistical properties of $\xi\bkt{t}$ as in Eq.\,\eqref{ximom}. We can see that averaging the above equation over the stochastic force ($\avg{...}_{\xi}$) reduces to Eq.(\ref{EffEqCOMAvg}), which we can interpret as the Langevin equation of motion after averaging over the possible noise realizations. As has been shown previously in \cite{CRV}, the correlations of the system observables can be obtained from the solutions of such a Langevin equation. 
\subsection{Generalized fluctuation-dissipation relation}\label{GFDR}
We now analyze  the relations between the dissipation and noise kernels of the composite environment. Considering the definitions of the dissipation and noise kernels given in Eqs.\eqref{Eta2} and \eqref{Nu2}, respectively, we first take the long-time limit. In this limit, the dissipation and noise kernels simplify to
\eqn{
\eta^{(2)} (t_1,t_2)\longrightarrow& \sbkt{\frac{1}{2}\eta ^{(+)} \bkt{t_1 - t_2} \nu_{GG} \bkt{ t_1,t_2}+\frac{1}{4}g\bkt{t_1 - t_2}\nu ^{(+)} \bkt{t_1 - t_2} }\Theta\bkt{t_1 - t_2},\label{Eta2LongTime}\\
\nu^{(2)} (t_1,t_2) \longrightarrow&  \frac{1}{2}\nu ^{(+)} \bkt{t_1 - t_2} \nu_{GG} \bkt{ t_1,t_2}-\frac{1}{4}g\bkt{t_1 - t_2} \eta ^{(+)} \bkt{t_1 - t_2},\label{Nu2LongTime}
}
where the term $\aavg{Q_h\bkt{ t_1} Q_h\bkt{t_2}}$ associated with the relaxation of the IDF vanishes in the late time limit, and the kernel $\nu_{GG} (t_1, t_2) $ reads:
\eqn{&\nu_{GG}(t_1,t_2)\equiv \int_{0}^{\infty}\dd\tau_1\int_{0}^{\infty}\dd\tau_2  {G}\bkt{t_1-\tau_1}\nu^{(-)}\bkt{\tau_1-\tau_2} {G}\bkt{t_2-\tau_2}.
\label{NuGGLongTime}
}
Notice that this limit holds regardless of the initial state of the IDF. This is in agreement with the notion that the dynamics at late-times is determined by the field fluctuations only.

It can be shown that the  Fourier and Laplace transforms of the dissipation and noise kernels corresponding to the individual baths can be related in terms of the following fluctuation-dissipation relations (see Appendix\,\ref{App:kernels} for details) %
\eqn{\label{FDRBathPM}\bar \nu^{(\pm)}(\omega)=&-2\coth\left(\frac{\omega}{2k_{B}T_{F}}\right){\rm Im}\sbkt{\bar\eta^{(\pm)}(\omega)}
\\
\label{FDRnuGG}
\bar \nu_{GG}(\omega)=&\coth\bkt{\frac{\omega}{2k_{B}T_{F}}}{\rm Im}[\bar{G
}(\omega)],
}
where $\bar F(\omega)$ represent the Fourier transforms of $F(t)$, defined  as  $\bar F(\omega)=\int_{-\infty}^{+\infty}{dt} e^{i\omega t} F(t)$. Note that both $\eta^{(\pm)}$ and $G$ are causal functions, as is also true for the dissipation kernel $\eta^{(2)}$.

%
One can thus rewrite Eqs.\eqref{Eta2LongTime} and \eqref{Nu2LongTime} in terms of the Fourier transformed kernels as follows
\eqn{\bar{\eta}^{(2)}(\omega)=&\frac{1}{2}\sbkt{\bar\nu_{GG}\ast\bar\eta^{(+)}}(\omega)+\frac{1}{4}\sbkt{\bar{G}\ast\bar{\nu}^{(+)}}(\omega),\label{Eta2W}\\
\bar\nu^{(2)}(\omega)=&\frac{1}{2}\sbkt{\bar\nu_{GG}\ast\bar\nu^{(+)}}(\omega)+\sbkt{{\rm Im}\bkt{\bar{G}}\ast{\rm Im}\bkt{\bar\eta^{(+)}}}(\omega).
\label{Nu2W}
}
 
The convolution product is defined for two functions $[A\ast B](\omega)\equiv\int_{-\infty}^{+\infty}\frac{d\omega'}{2\pi}A(\omega-\omega')B(\omega')$, which satisfies $[A\ast B](\omega)=[B\ast A](\omega)$.

Using the fluctuation-dissipation relations for the kernels of the two baths and the IDF, we can write the  dissipation kernel for the MDF in terms of those of the IDF and the baths in frequency domain as follows  
%
%
%
\eqn{ \label{eta2w2}{\rm Im}\sbkt{\bar{\eta}^{(2)}(\omega)}=&\int_{0}^{+\infty}\frac{d\omega'}{2\pi}\frac{1}{2}\left[\frac{1}{2}\cbkt{\bar\nu^{(+)}(\omega-\omega')-\bar \nu^{(+)}(\omega+\omega')}{\rm Im}[\bar{G}(\omega')]+\cbkt{\bar{\nu}_{GG}(\omega-\omega')-\bar{\nu}_{GG}(\omega+\omega')}{\rm Im}\sbkt{\bar \eta^{(+)}(\omega')}\right].
}

We note from the above equation that the contributions to the total dissipation come from the three body processes between the MDF, IDF and the $ (+) $ bath, with the propagator of the IDF combining with the noise of the $(+)$ bath and vice versa.  Each term in the above equation accounts for the four possible processes involved in the dynamics at second order, wherein one of the thermal baths contributes the initial excitation, while the other bath acts as an extra channel for partial energy exchange.

Furthermore, the above kernel can be alternatively expressed in a compact way as follows \eqn{
{\rm Im}\sbkt{\bar{\eta}^{(2)}(\omega)}=&\int_{-\infty}^{+\infty}\frac{d\omega'}{2\pi}\frac{1}{2}\left[\coth\left(\frac{\omega-\omega'}{2k_{B}T_{F}}\right)-\coth\left(\frac{\omega'}{2k_{B}T_{F}}\right)\right]{\rm Im}[\bar {G}(\omega-\omega')]~{\rm Im}\sbkt{\bar \eta^{(+)}(\omega')}.
\label{ImEta2DissTimesNoise}}
%

We can similarly express the noise kernel in terms of the dissipation kernels of the IDF and the two baths as follows
%
%
\eqn{\label{nu2w2}\bar{\nu}^{(2)}(\omega)=&\int_{-\infty}^{+\infty}\frac{d \omega'}{2\pi}\frac{1}{2}\sbkt{1-\coth\bkt{\frac{\omega-\omega'}{2k_{B}T_{F}}}\coth\bkt{\frac{\omega'}{2k_{B}T_{F}}}}{\rm Im}[\bar {G}(\omega-\omega')]~{\rm Im}\sbkt{\bar \eta^{(+)}(\omega')}.
}
It is not possible to find simple relation between the above tranforms of the dissipation and noise kernels $\{\bar{\eta}^{(2)},\bar{\nu}^{(2)}\}$ because of the presence of the integrals on $\omega'$, which are related to the convolution products.  We therefore define  new kernels  $D^{(2)}(\omega,\omega')$ and $N^{(2)}(\omega,\omega')$   that depend on two frequency variables,  accounting for processes where a given frequency-component contribution is modified by other frequencies such that $\bar\eta^{(2)}(\omega)=\int_{-\infty}^{+\infty}\frac{d\omega'}{2\pi}D^{(2)}(\omega,\omega')$ and  $\bar\nu^{(2)}(\omega)=\int_{-\infty}^{+\infty}\frac{d\omega'}{2\pi}N^{(2)}(\omega,\omega')$.  Thus we can find a relation between these new kernels that reads
\eqn{N^{(2)}\bkt{\omega,\omega'}=2\coth\sbkt{\frac{\omega-2\omega'}{2k_{B}T}}{\rm Im}\sbkt{D^{(2)}(\omega,\omega')},
\label{FDRImEta2ReNu2}
}
or inversely
\eqn{{\rm Im}\sbkt{D^{(2)}(\omega,\omega')}=\frac{1}{2}\tanh\sbkt{\frac{\omega-2\omega'}{2k_{B}T}}N^{(2)}\bkt{\omega,\omega'}.
\label{FDRImEta2ReNu2bis}
}

The introduction of an extra variable $\omega'$ stands for the complexity of the environment in terms of its energy exchange with the system. The nonlinear coupling allows the two parts of the environment (the field and the IDF) to simultaneously exchange energy with the system. From these relations, it becomes simple to prove that
\eqn{{\rm Im}\sbkt{\bar \eta^{(2)}(\omega)}
=&\int_{-\infty}^{+\infty}\frac{d\omega'}{2\pi}\frac{1}{2}\tanh\sbkt{\frac{\omega'}{2k_{B}T}}N^{(2)}\bkt{\omega,\frac{\omega-\omega'}{2}},
\label{DissAsIntN2DoubleW}
}
and
\eqn{\bar\nu^{(2)}(\omega)
=&\int_{-\infty}^{+\infty}\frac{d\omega'}{2\pi}2\coth\sbkt{\frac{\omega'}{2k_{B}T}}D^{(2)}\bkt{\omega,\frac{\omega-\omega'}{2}}.
\label{NoiseAsIntD2DoubleW}
}
%
The above equation corresponds to a generalized  fluctuation-dissipation relation (FDR) for the late-time dissipation and noise kernels.

We can physically interpret the results by  comparing the present situation  to the case of linear QBM.  The FDR that holds between the dissipation and noise kernels in QBM can be stated in frequency-domain for a single frequency component $\omega$ in the general form that reads $[{\textit Noise}](\omega)=[{\textit Thermal Factor}](\omega)\times[{\textit Dissipation}](\omega)$, wherein a single frequency connects both sides of the relation.  In our case, the nonlinear nature of the  coupling precludes such a relation at a single frequency.  Physically such a nonlinear coupling leads  to processes wherein an excitation from one of the baths interacts with the system while perturbed by the other bath. We can thus formulate a generalized FDR Eq.\,\eqref{FDRImEta2ReNu2} by introducing two frequency variables to account for the nonlinearity.

\section{Discussion}
\label{Discussion}

We have derived the non-equilbrium dissipative center of mass dynamics of a particle interacting with a field via its IDF. The model considered in this paper underlies  microscopic optomechanical interactions between neutral particles and fields, and is generally applicable to open quantum systems that possess intermediary quantum degrees of freedom between system and bath.    We show that the field can be separated into two baths,  referred to as $(+)$ and $(-)$ baths, with the $(-)$ bath coupled linearly to the IDF, and the $(+)$ bath coupled nonlinearly to both IDF and MDF (see Eq.\,\eqref{Sint-} and  Eq.\,\eqref{Sint+}).  Such a decomposition allows one to systematically trace over $(-)$ bath and express its resulting influence  on the IDF in terms of a exact second order effective action as given by Eq.\,\eqref{smineff}. As the nonlinear coupling between the MDF, IDF and the $(+)$ bath poses a challenge in terms of writing the exact dynamics of the MDF, we use a perturbative influence functional approach to trace over the IDF and $(+)$ bath to obtain a second order effective action for the MDF. We find that the  effective action, the dissipation and noise kernels resulting from the composite environment (Eq.\,\eqref{EffAct2nd}, \eqref{Eta2} and \eqref{Nu2} respectively) are  structurally similar  to linear QBM dynamics. This can be attributed to the quadratic nature of coupling between the system variable  and the composite bath that goes as $\sim X\mc{B}$, where the bath operator $\mc{B}\bkt{ \sim Q \sum_n q_n ^{(+)}}$ can be nonlinear. We find a Langevin equation of motion for the MDF, describing its dissipative non-equilibrium dynamics (Eq.\,\eqref{EffEqCOMLangevin}). The dissipation and noise kernels at late time can be related via a generalized FDR as given by Eq.\,\eqref{FDRImEta2ReNu2}.

 This work highlights three aspects:
\begin{enumerate}
    \item {\textit{Dissipation and noise of the open quantum system in the presence of an IDF} -- It can be seen from the  dissipation and noise kernels  (Eq.\,\eqref{Eta2} and \eqref{Nu2}) arising from the composite bath of the IDF+field that the  total dissipation of the MDF is a combination of the dissipation kernel corresponding to the IDF and noise kernel corresponding to the $(+) $ bath, and vice versa. Similarly the noise kernel involves a combination of the noise of the IDF and that of the  $(+)$ bath. Additionally, it also includes a contribution resulting from the combination of the dissipation of the IDF and that of the $(+)$ bath, which is needed to obtain a generalized FDR for the kernels of our composite environment. As discussed  in  Sec.\,\ref{TracingInternalDOF}, the dissipation and noise kernels are similar in structure to those of linear QBM up to the second order perturbative in action. The present approach can be extended to higher orders in a systematic way, thus  extending the results of Ref.\cite{HPZ93} to complex environments.}
    
    \item{ \textit{Coupled non-equilibrium dynamics of different degrees of freedom with self-consistent backaction} -- The equation of motion for the MDF given in Eq.\,\eqref{EffEqCOMLangevin} describes the quantum center of mass motion including the self-consistent backaction of the various degrees of freedom on each other. The approach based on the separation of the field into two uncorrelated baths is akin to the Einstein-Hopf theorem that establishes the statistical independence of the blackbody radiation field and its derivative \cite{EinHopf1,  EinHopf3}. The influence of the $(-)$ bath on the IDF is included via the effective action Eq.\,\eqref{smineff}, and the perturbative second order effective action in Eq.\,\eqref{EffAct2nd} includes the influence of the $(+)$ bath and the IDF (affected by the $(-)$ bath). In this way, the quantum field influences the MDF's dynamics by two means, by its direct interaction as the $(+)$ bath and through the fluctuations of the IDF as the $(-)$ bath.  Furthermore, considering the zero-temperature limit of the dynamics, we find that  a mechanical oscillator interacting with the vacuum field exhibits dissipative dynamics as a result of the quantum fluctuations of its composite nonlinear environment. Such a vacuum-induced noise poses a fundamental constraint on preparing mechanical objects in quantum states \cite{KSYS1, Skatepark}.}
    
    \item{\textit{Fluctuation-dissipation relations for a system coupled to a composite environment } -- We find a generalized FDR between the late-time dissipation and noise kernels, that has a similar structure to linear QBM (see Eq.\,\eqref{FDRImEta2ReNu2}) albeit involving two frequency variables due to the nonlinearity of the interaction. This allows one to interpret the single-frequency response of the effective kernels  for the MDF in terms of contributions from various frequency  components of the IDF and field. Our result provides an example of FDR for an  open quantum system  with nonlinear dynamics when linear response theory is no longer valid \cite{HPZ93, JT20}.}
\end{enumerate} 
As a future prospect our results can be extended to studying the non-equilibrium dynamics in a wide range of physical systems, particularly when the time scales of the different degrees of freedom are no longer disparate. In the context of atom-field interactions, it has been shown for example that the internal and external degrees of freedom of atoms in a standing laser wave can exhibit synchronization  \cite{Argonov05}. Furthermore, it is possible for  the IDF dynamics to be sufficiently slow for a dark internal state which can lead to highly correlated internal-external dynamics as in the case of velocity selective coherent population trapping scheme \cite{Aspect88, Aspect89},  and  long-lived internal states in atomic clock transitions \cite{Boyd}. It has also been experimentally observed that the internal and external degrees of freedom can be efficiently coupled in the presence of colored noise \cite{Machluf10}. On the other hand, when the center of mass dynamics is fast enough to be comparable to the internal dynamics, one requires a careful consideration of the coupled internal-external dynamics. Such a situation can arise in the case of dynamical Casimir effect (DCE) \cite{Lin18}, when considering the self-consistent backaction of the field and the IDF on the mirror \cite{Butera, Belen19}. Finally,  our results can be extended to analyze the thermalization properties and non-equilibrium dynamics of the center mass of  nanoparticles \cite{Yin, AERL18}.

\section{Acknowledgments}
We are grateful to Bei-Lok Hu, Esteban A. Calzetta,  and Peter W. Milonni  for insightful discussions.
Y.S. acknowledges support from the Los Alamos National Laboratory ASC Beyond Moore's Law project and the LDRD program. A.E.R.L. wants to thank the AMS for the support.

\appendix

\section{Generating functional derivation}
\label{App:gf}
 The generating functionals corresponding to the IDF and the $(+) $ bath can be explicitly written as
\eqn{\label{gfidf}
\mc{F}^{(1)}_I\sbkt{J,J'} = \int\dd Q_f\int\dd Q_i\int\dd Q'_i\,\rho_I\bkt{Q_i,Q'_i;0}\int_{Q(0) = Q_i}^{Q(t) = Q_f}\ddd Q\int_{Q'(0) = Q'_i}^{Q'(t) = Q_f}\ddd Q'& e^{i \bkt{S_I[Q]-S_I[Q'] +S_{\mr{I, IF}}^{(-)}[Q,Q']}}\non\\
&e^{i\int_0^t \dd\tau \sbkt{J(\tau)Q(\tau)-J'(\tau)Q'(\tau)}}
}
\eqn{\label{gfqn}
\mc{F}^{(1)}_n\sbkt{J_n,J'_n} =&\int\dd q_{nf}^{(+)}\int\dd q_{ni}^{(+)}\int\dd {q_{ni}^{(+)}}'\,\rho_{F,n}^{(+)}\bkt{\cbkt{q_{ni}^{(+)},{q_{ni}^{(+)}}'};0}\non\\
&\int_{q_n^{(+)}(0) = q^{(+)}_{ni}}^{q^{(+)}_{n}(t) = q^{(+)}_{nf}}\ddd q^{(+)}_n\int_{{q_n^{(+)}}'(0) = {q_{ni}^{(+)}}'}^{{q_n^{(+)}}'(t) = q^{(+)}_{nf}}\ddd {{q_n}^{(+)}}' e^{i \bkt{S_{F,n}^{(+)}[\cbkt{q_n^{(+)}}]-S_{F,n}^{(+)}[\cbkt{{q_n^{(+)}}'}] }}
e^{i\int_0^t \dd\tau \sbkt{J_n(\tau)q_n^{(+)}(\tau)-J'_n(\tau){{q_n^{(+)}}'}(\tau)}},
}
where $\rho_{F,n}^{(+)}\bkt{q_{ni}^{(+)},{q_{ni}^{(+)}}';0}$ represents the initial state of the $n^\mr{th}$ oscillator of the $(+) $ bath, and $S_{F,n}^{(+)}$ is the corresponding free action. We can then see from Eq.\,\eqref{IF-} and Eq.\,\eqref{Fexp-} that the generating functional for the $(+)$ bath  is simply the influence functional for linear QBM given by
\eqn{\label{gf1}\mc{F}^{(1)}_n\sbkt{J_n,J'_n} = e^{i S_{\mr{IF},n}^{(+)}[J_n,J_n']},
}

where influence action $S_{\mr{IF},n} ^{(+)}[J_n,J_n']$ is defined as

\eqn{ \label{sif+}
S^{(+)}_{\mr{IF},n}\sbkt{  J_n, J_n'}\equiv-\int_0^t\dd \tau_1\int_0^{\tau_1}\dd\tau_2 \sbkt{  J_n\bkt{\tau_1} -  J_n'\bkt{\tau_1} }\eta_n^{(+)}\bkt{\tau_1 - \tau_2} \sbkt{  J_n \bkt{\tau_2} +  J_n' \bkt{\tau_2}}\non\\
+i\int_0^t\dd \tau_1\int_0^{\tau_1}\dd\tau_2 \sbkt{  J_n\bkt{\tau_1} -  J_n'\bkt{\tau_1} }\nu_n^{(+)}\bkt{\tau_1 - \tau_2} \sbkt{  J_n \bkt{\tau_2} -  J_n' \bkt{\tau_2}},
}

with the dissipation and noise kernels

\eqn{\label{eta+}
\eta_n^{(+)}(\tau) =&-  \frac{1 }{2 \omega_n }\sin\bkt{\omega_n \tau}\\
\label{nu+}
\nu_n^{(+)}(\tau) =&  \frac{1}{2 \omega_n }\coth\bkt{\frac{ \omega_{n}}{2k_BT_F}}\cos\bkt{\omega_n \tau}.
}

Now to evaluate the generating functional for the IDF in terms of $\cbkt{J,J'}$, we follow the approach in \cite{CRV}. Consider the effective action pertaining to the IDF in Eq. \eqref{gfidf}
\begin{widetext}
\eqn{
S_{I,\mr{eff}}\sbkt{Q,Q',J,J'} =& S_I\sbkt{Q}-S_I\sbkt{Q'}+S_{I,\mr{IF}}^{(-)}\sbkt{Q,Q'}+\int_0^t\dd \tau \sbkt{J(\tau)Q(\tau)-J'(\tau)Q'(\tau)}\\
 =& \int_0^t \dd \tau \sbkt{\frac{1}{2} M\dot Q^2 - \frac{1}{2} M\omega_0^2Q^2 +J Q} - \sbkt{\frac{1}{2} M\dot Q'^2 - \frac{1}{2} M\omega_0^2Q'^2 +J' Q'} \non\\
&- \int _0^t \dd\tau_1\int _0^{\tau_1} \dd\tau_2 \sbkt{Q(\tau_1) -Q'(\tau_1)  }\eta^{(-)}\bkt{\tau_1-\tau_2}\sbkt{Q(\tau_2) +Q'(\tau_2)  }\non\\
&+i \int _0^t \dd\tau_1\int _0^{\tau_1} \dd\tau_2 \sbkt{Q(\tau_1) -Q'(\tau_1)  }\nu^{(-)}\bkt{\tau_1-\tau_2}\sbkt{Q(\tau_2) -Q'(\tau_2)  }\\
 =& \int_0^t \dd \tau \sbkt{ \frac{1}{2} M\dot Q_\Sigma(\tau)\dot Q_\Delta(\tau) - \frac{1}{2}M\omega_0^2 Q_\Sigma(\tau)Q_\Delta(\tau) +Q_\Sigma(\tau) J_\Delta(\tau)+ Q_\Delta(\tau) J_\Sigma(\tau)} \non\\
&- \int _0^t \dd\tau_1\int _0^{\tau_1} \dd\tau_2 Q_\Delta(\tau_1)\eta^{(-)}\bkt{\tau_1-\tau_2}Q_\Sigma(\tau_2)+i \int _0^t \dd\tau_1\int _0^{\tau_1} \dd\tau_2 Q_\Delta(\tau_1)\nu^{(-)}\bkt{\tau_1-\tau_2}Q_\Delta(\tau_2),
}
\end{widetext}
where we have defined the new  coordinates as $Q_\Sigma \equiv Q+Q'$, $Q_\Delta\equiv Q-Q'$, $J_\Sigma\equiv \bkt{J+J'}/2$, and $J_\Delta\equiv \bkt{J-J'}/2$. The dissipation and noise kernels $\eta^{(-)} \bkt{t} $ and $\nu^{(-)}\bkt{t}$ are as defined in Eq.\,\eqref{DissQBMMinus} and Eq.\,\eqref{NoiseQBMMinus}.

Rewriting the generating functional  in terms of the relative and center of mass coordinates, using the approach outlined in \cite{CRV}, Eq.\,\eqref{gfidf} can be simplified to obtain

\eqn{\label{gfidf2}
\mc{F}^{(1)}_I\sbkt{J,J'} 
=& \aavg{e^{2iJ_\Delta\cdot \,Q_{h}}} e^{-2\bkt{J_\Delta\cdot\,  {G}} \cdot {\nu}^{(-)}\cdot\bkt{J_\Delta\cdot \,{G}}^T}e^{2iJ_\Delta\cdot {G}\cdot J_\Sigma},
}
where \eqn{\label{aavg} \aavg{\dots} = \int \dd Q^i \dd \Pi^i W\bkt{Q^i, \Pi^i}( \dots ),} denotes the average over the initial Wigner distribution $W(Q^i, \Pi^i)$.  $Q_{ h}$ is the solution to the homogeneous Langevin equation \eqn{\label{IDFhom}{L}\cdot Q_h =0,} with initial conditions given by $\cbkt{Q^i,\Pi^i}$ and the differential operator $L(t,t')=m(\frac{d^{2}}{dt'^{2}}+\omega_{0}^{2})\delta(t-t')+2\eta^{(-)}(t-t')$.  The Green's function corresponding to the Langevin operator ${{L}}$ is defined as \eqn{\label{Gret}{ L}\cdot { G} = { \delta},}
where $ A\cdot B\equiv \int_0 ^t \dd \tau A(\tau ) B(\tau) $.

\section{Perturbative effective action derivation}
\label{App:seff}
Now having obtained the total generating functional as defined in Eq. \eqref{Gentot} which contains the influence of the IDF  (Eq. \eqref{gfidf2}) and the (+) bath (Eq. \eqref{sif+}) on the system,  we substitute and back in the total generating functional  to obtain the effective action  perturbatively up to second order in the system-bath coupling parameter in a term by term manner as follows.

\eqn{F[X,X'] =& \left.\exp{\cbkt{{i}\bkt{\sint^{(+)} \sbkt{X,\frac{1}{i}\frac{\delta}{\delta J},\cbkt{\frac{1}{i}\frac{\delta}{\delta J_n}}} - \sint^{(+)} \sbkt{X',\frac{-1}{i}\frac{\delta}{\delta J'},\cbkt{\frac{-1}{i}\frac{\delta}{\delta J'_n} }}}}}\mc G^{(1)}[ J, J',\cbkt{J_n, J'_n}]\right\vert_{{\bf J} = {\bf J'}=0}\non\\
&\equiv \exp{ \cbkt{{i} \seff [X,X']}}.
}
So far the above expression would yield the exact effective action for the given system-bath interaction since we have not invoked any weak-coupling approximations yet. We will now make a perturbative expansion of the interaction action up to second order in the system-bath coupling strength in the exponent. Let us consider $\sint^{(+)} \sbkt{X,\frac{1}{i}\frac{\delta}{\delta J},\cbkt{\frac{1}{i}\frac{\delta}{\delta J_n}}} \equiv \varepsilon \tilde {S}_{\mr{int}}^{(+)}\sbkt{X,\frac{1}{i}\frac{\delta}{\delta J},\cbkt{\frac{1}{i}\frac{\delta}{\delta J_n}}}$ and $\sint^{(+)} \sbkt{X',-\frac{1}{i}\frac{\delta}{\delta J'},\cbkt{-\frac{1}{i}\frac{\delta}{\delta J'_n}}} \equiv \varepsilon' \tilde {S}_{\mr{int}}^{(+)}\sbkt{X',-\frac{1}{i}\frac{\delta}{\delta J'},\cbkt{-\frac{1}{i}\frac{\delta}{\delta J'_n}}}$, where the dimensionless parameters $\varepsilon$ and $\varepsilon'$ characterize the system-bath coupling strength that we expand the influence action about
\eqn{\seff^{(2)}[X,X'] \approx \left.\seff[X,X']\right\vert_{\varepsilon= \varepsilon'=0}+ \varepsilon\left.\frac{\delta}{\delta\varepsilon}\seff[X,X']\right\vert_{ \varepsilon =  \varepsilon'=0}+  \varepsilon'\left.\frac{\delta}{\delta \varepsilon'}\seff[X,X']\right\vert_{ \varepsilon =  \varepsilon'=0}\nonumber\\+\frac{ \varepsilon^2}{2} \left.\frac{\delta^2}{\delta \varepsilon^2}\seff[X,X']\right\vert_{ \varepsilon =  \varepsilon'=0}
+\frac{ \varepsilon'^2}{2} \left.\frac{\delta^2}{\delta \varepsilon'^2}\seff[X,X']\right\vert_{ \varepsilon =  \varepsilon'=0}+{ \varepsilon \varepsilon'}\left.\frac{\delta^2}{\delta \varepsilon\delta \varepsilon'}\seff[X,X']\right\vert_{ \varepsilon =  \varepsilon'=0}
}
Let us consider the above expression term by term

\begin{enumerate}
\item{$\left.\seff[X,X']\right\vert_{ \varepsilon =  \varepsilon'=0} = 0$, this can be understood as the influence action corresponding to a non-interacting bath, which trivially vanishes.}
\item{$\begin{aligned}[t]
\left.\frac{\delta}{\delta \varepsilon}\seff[X,X']\right\vert_{ \varepsilon =  \varepsilon'=0} &=\left. \tilde {S}_\mr{int}^{(+)}\sbkt{X,\frac{1}{i}\frac{\delta}{\delta{\bf J}}}\mc G^{(1)}[\bf{J},\bf{J'}]\right\vert_{{\bf J} = {\bf J'}=0}
 \end{aligned}
$
}
\item{$\begin{aligned}[t]
 \left.\frac{\delta}{\delta \varepsilon'}\seff[X,X']\right\vert_{ \varepsilon =  \varepsilon'=0} &= -\left. \tilde {S}_\mr{int}^{(+)}\sbkt{X',-\frac{1}{i}\frac{\delta}{\delta{\bf J'}}}\mc G^{(1)}[\bf{J},\bf{J'}]\right\vert_{{\bf J} = {\bf J'}=0}
 \end{aligned}$}
 \item{$\begin{aligned}[t]
  \left.\frac{\delta^2}{\delta \varepsilon^2}\seff[X,X']\right\vert_{ \varepsilon =  \varepsilon'=0} ={i}\bkt{ \left.\sbkt{\bkt{\tilde{S}_\mr{int}^{(+)}\sbkt{X,\frac{1}{i} \frac{\delta}{\delta  {\bf J}}}}^2\mc G^{(1)}[\bf{J},\bf{J'}]}\right\vert_{{\bf J} = {\bf J'}=0}- \sbkt{\left.\tilde{S}_\mr{int}^{(+)}\sbkt{X,\frac{1}{i} \frac{\delta}{\delta  {\bf J}}}\mc G^{(1)}[\bf{J},\bf{J'}]\right\vert_{{\bf J} = {\bf J'}=0}}^2}
 \end{aligned}$}
\item{$\begin{aligned}[t]
 \left.\frac{\delta^2}{\delta \varepsilon'^2}\seff[X,X']\right\vert_{ \varepsilon =  \varepsilon'=0} = i&\left( \left.\sbkt{\bkt{\tilde{S}_\mr{int}^{(+)}\sbkt{X',-\frac{1}{i} \frac{\delta}{\delta  {\bf J'}}}}^2\mc G^{(1)}[\bf{J},\bf{J'}]}\right\vert_{{\bf J} = {\bf J'}=0}\right.\nonumber\\
 &\left.-  \sbkt{\left.\tilde{S}_\mr{int}^{(+)}\sbkt{X',-\frac{1}{i} \frac{\delta}{\delta  {\bf J'}}}\mc G^{(1)}[\bf{J},\bf{J'}]\right\vert_{{\bf J} = {\bf J'}=0}}^2\right)
 \end{aligned}$ }
\item{$\begin{aligned}[t]
\left.\frac{\delta^2}{\delta \varepsilon\delta \varepsilon'}\seff[X,X']\right\vert_{ \varepsilon =  \varepsilon'=0} =&-i\left( \left.\sbkt{\tilde{S}_\mr{int}^{(+)}\sbkt{X,\frac{1}{i} \frac{\delta}{\delta  {\bf J}}}\tilde{S}_\mr{int}^{(+)}\sbkt{X',-\frac{1}{i} \frac{\delta}{\delta  {\bf J'}}}\mc G^{(1)}[\bf{J},\bf{J'}]}\right\vert_{{\bf J} = {\bf J'}=0}\right.\nonumber\\
&\left.-  \sbkt{\left.\tilde{S}_\mr{int}^{(+)}\sbkt{X,\frac{1}{i} \frac{\delta}{\delta  {\bf J}}}\mc G^{(1)}[\bf{J},\bf{J'}]\right\vert_{{\bf J} = {\bf J'}=0}}\sbkt{\left.\tilde{S}_\mr{int}^{(+)}\sbkt{X',-\frac{1}{i} \frac{\delta}{\delta  {\bf J'}}}\mc G^{(1)}[\bf{J},\bf{J'}]\right\vert_{{\bf J} = {\bf J'}=0}}\right)
\end{aligned}$
 }
\end{enumerate}
Putting all the terms together we can rewrite the influence action up to second order in the system-bath coupling as in Eq\,\eqref{seff}.
\section{Calculation of averages}
\label{Seffavg}
\subsection{First order average}

The first order average in the influence action can be calculated as follows
\eqn{\avg{\sint^{(+)}\sbkt{{X},\frac{1}{i}\frac{\delta}{\delta { J}},\cbkt{\frac{1}{i}\frac{\delta}{\delta { J_n}}}}}_0 &= \left. \sint^{(+)}\sbkt{{X},\frac{1}{i}\frac{\delta}{\delta { J}},\cbkt{\frac{1}{i}\frac{\delta}{\delta { J_n}}}}\mc G^{(1)}[{J},{J'},\cbkt{J_n,J_n'}]\right\vert_{{\bf J} = {\bf J'}=0}\nonumber\\
& =  \left. \int_0^t \dd \tau \sum_{n}\lambda \omega_{n}\sbkt{\frac{1}{i}\frac{\delta}{\delta{ J_n(\tau)}}}\sbkt{ \frac{1}{i}\frac{\delta}{\delta{J(\tau)}}} {X}(\tau)\prod_{p} F^{(1)}_p [J_p,J_p']F^{(1)}_I [J,J']\right\vert_{{\bf J} = {\bf J'}=0}
}
where we have used the  generating functional  as defined in Eq. \eqref{Gentot}, and the interaction action from Eq. \eqref{Sint+}. Now one can note that the first order derivative of the influence action which is quadratic in $ \cbkt{J_n,J_n'} $ brings in a factor that is linear in $\cbkt{J_n,J_n'}$, and thus vanishes when we set $J_n = J_n'=0$. Thus, we obtain that
\eqn{&\avg{\sint^{(+)}\sbkt{X,\frac{1}{i}\frac{\delta}{\delta { J}},\cbkt{\frac{1}{i}\frac{\delta}{\delta { J_n}}}}}_0=\avg{\sint^{(+)}\sbkt{X',\frac{1}{i}\frac{\delta}{\delta { J'}},\cbkt{\frac{1}{i}\frac{\delta}{\delta { J'_n}}}}}_0 =0.}

\subsection{Second order averages}
\label{AppB2}
We calculate the second order terms in the influence action as follows:

\begin{enumerate}
    \item {
\begin{widetext}
\eqn{\label{seffxx}
\avg{\cbkt{\sint^{(+)}\sbkt{X,\frac{1}{i}\frac{\delta}{\delta {\bf J}}}}^2}_0  =&   \bkt{\int_0^t \dd t_1 \sum_{n}\lambda \omega_{n}\sbkt{\frac{1}{i}\frac{\delta}{\delta{ J(t_1)}}}\sbkt{ \frac{1}{i}\frac{\delta}{\delta{J_{n}(t_1)}}} X(t_1)}\nonumber\\
&\bkt{\int_0^t \dd t_2 \sum_{m}\lambda k_{m}\sbkt{\frac{1}{i}\frac{\delta}{\delta{ J{}(t_2)}}}\sbkt{ \frac{1}{i}\frac{\delta}{\delta{J_{m}(t_2)}}} X(t_2)}\left.F^{(1)}_I [J,J']\,\prod_{p}F^{(1)}_p [J_p,J_p']\right\vert_{{\bf J}
= {\bf J'}=0}\\
\label{seffxx2}
&= \int_0^t \dd t_1\int_0^t \dd t_2 \sum_{m,n}\lambda^2 \omega_{m}\omega_{n} X(t_1) X(t_2)\,\zeta_I\bkt{t_1,t_2}\,\sum_{p}\zeta^{(p)}_{m,n}\bkt{t_1,t_2}
}
\end{widetext}
where we have defined the influence of the  (+) bath and the IDF as 

\eqn{
\label{zetap}\zeta^{(p)}_{m,n}(t_1,t_2)\equiv &\left.\frac{\delta }{\delta J_n(t_1)}\cbkt{\frac{\delta }{\delta J_m(t_2)}\sbkt{F^{(1)}_p [J_p,J'_p]}}\right\vert_{J_p = J'_p =0}\non\\
=& -{i}\delta_{mp}\delta_{np}\sbkt{\eta_p^{(+)} \bkt{t_1-t_2 }\Theta\bkt{t_1-t_2}+\eta_p^{(+)}\bkt{t_2-t_1}\Theta\bkt{t_2-t_1}}\non\\
&-\delta_{mp}\delta_{np}\sbkt{\nu_p^{(+)} \bkt{t_1-t_2}\Theta\bkt{t_1-t_2}+\nu_p^{(+)}\bkt{t_2-t_1}\Theta\bkt{t_2-t_1}}\\
= & \delta_{mp}\delta_{np}\sbkt{-i\eta_p^{(+)} \bkt{t_1-t_2 }\mr{sign}\bkt{t_1-t_2}-\nu_p^{(+)}\bkt{t_1-t_2}}}
\eqn{
\label{zetaI}
\zeta_I(t_1,t_2)\equiv& \frac{\delta }{\delta J(t_1)}\cbkt{\frac{\delta }{\delta J(t_2)}\sbkt{F^{(1)}_I [J,J']}}\non\\
 =& -\frac{1}{2} \sbkt{\int_0^{t} \dd \tau_1 \int_0^t \dd \tau_2\bkt{ {G} \bkt{t_1-\tau_1} \nu^{(-)}\bkt{\tau_1-\tau_2}   {G}\bkt{t_2 -\tau_2} +   {G}\bkt{t_2 - \tau_1}\nu^{(-)}\bkt{\tau_1-\tau_2}   {G}\bkt{t_1 - \tau_2}}}\non\\
&-\frac{i}{2} \sbkt{  {G}\bkt{t_1-t_2}+  {G}\bkt{t_2-t_1}} - \aavg{ Q_h\bkt{ t_1} Q_h\bkt{t_2}}.
}
Let us further define an odd function $ {g}\bkt{t}$ such that $  {G}\bkt{t_1 - t_2} \equiv    {g}\bkt{t_1 - t_2}\Theta \bkt{t_1 - t_2}$, and \eqn{\nu_{GG}\bkt{t_1 , t_2}\equiv &\int_0^{t_1} \dd \tau_1 \int_0^{t_2} \dd \tau_2\sbkt{   {g} \bkt{t_1-\tau_1} \nu^{(-)}\bkt{\tau_1-\tau_2} {g}\bkt{t_2 -\tau_2}},
}
such that we can rewrite $\zeta_I \bkt{t_1, t_2}$ as
\eqn{ \zeta_I \bkt{t_1, t_2 } = - \nu_{GG}\bkt{t_1 , t_2} - \frac{i }{2}  {g} \bkt{t_1 - t_2} \mr{sign} \bkt{t_1 - t_2 }- \aavg{ Q_h\bkt{ t_1} Q_h\bkt{t_2}}.
}
We have made use of the fact that the function  $g\bkt{s} = -g\bkt{-s} $ is odd, and the noise kernel  $\nu ^{(+) }\bkt{s} =  \nu ^{(+) }\bkt{-s} $ is even.

This allows us to rewrite Eq.\eqref{seffxx2} as
\eqn{&\avg{\cbkt{\sint^{(+)}\sbkt{X,\frac{1}{i}\frac{\delta}{\delta {\bf J}}}}^2}_0\non\\
=& \int_0^t \dd t_1\int_0^t \dd t_2 \sum_p\lambda^2 k_p^2 X(t_1) X(t_2)\sbkt{-i \eta_p ^{(+) } \bkt{t_1 - t_2} \mr{sign}\bkt{t_1 - t_2 } - \nu_p ^{(+) } \bkt{t_1 - t_2 }} \non\\
&\sbkt{- \nu_{GG}'\bkt{t_1 , t_2} + \frac{i }{2}  {g} \bkt{t_1 - t_2} \mr{sign} \bkt{t_1 - t_2 }}\\
\label{seffXX}
= &\int_0^t \dd t_1\int_0^t \dd t_2 \sum_p\lambda^2 k_p^2 X(t_1) X(t_2)\sbkt{i \eta_p ^{(+)}\bkt{t_1 - t_2 } \nu_{GG}' \bkt{t_1, t_2}\mr{sign}\bkt{t_1-t_2}+ \nu_p ^{(+)} \bkt{t_1 - t_2} \nu_{GG}' \bkt{t_1, t_2} \right.\non\\
&\left. - \frac{1}{2} {g}\bkt{t_1 - t_2} \eta_p ^{(+)} \bkt{t_1 - t_2}+ \frac{i}{2} {g}\bkt{t_1 - t_2} \nu_p ^{(+)} \bkt{t_1 - t_2}\mr{sign}\bkt{t_1-t_2} },
}
where $\nu_{GG}' \bkt{ t_1,t_2} \equiv \nu_{GG} \bkt{ t_1,t_2} + \aavg{ Q_h\bkt{ t_1} Q_h\bkt{t_2}}$.
}
\item{
\eqn{ \label{seffxpxp}\avg{\cbkt{\sint^{(+)}\sbkt{X',-\frac{1}{i}\frac{\delta}{\delta {\bf J'}}}}^2}_0 =  \int_0^t\dd t_1 \int_0^t\dd t_2 \sum_{m,n} \lambda^2\omega_m\omega_n X'(t_1)X'(t_2)\,\widehat{\zeta}_I\bkt{t_1,t_2}\,\sum_{p}\widehat{\zeta}^{(p)}_{m,n}\bkt{t_1,t_2},
}
where
\eqn{\label{hatzetap}\widehat{\zeta}^{(p)}_{m,n}(t_1,t_2)\equiv &\left.\frac{\delta }{\delta J'_n(t_1)}\cbkt{\frac{\delta }{\delta J'_m(t_2)}\sbkt{F^{(1)}_p [J_p,J'_p]}}\right\vert_{J_p = J'_p =0}\non\\
=& {i}\delta_{mp}\delta_{np}\sbkt{\eta_p^{(+)} \bkt{t_1-t_2 }\Theta\bkt{t_1-t_2}+\eta_p^{(+)}\bkt{t_2-t_1}\Theta\bkt{t_2-t_1}}\non\\
&-\delta_{mp}\delta_{np}\sbkt{\nu_p^{(+)} \bkt{t_1-t_2}\Theta\bkt{t_1-t_2}+\nu_p^{(+)}\bkt{t_2-t_1}\Theta\bkt{t_2-t_1}} \\
= &  \delta_{mp}\delta_{np}\sbkt{i\eta_p^{(+)} \bkt{t_1-t_2 }\mr{sign}\bkt{t_1-t_2}-\nu_p^{(+)}\bkt{t_1-t_2}}
}
\eqn{
\label{hatzetaI}
\widehat{\zeta}_I(t_1,t_2)\equiv& \frac{\delta }{\delta J'(t_1)}\cbkt{\frac{\delta }{\delta J'(t_2)}\sbkt{F^{(1)}_I [J,J']}}\non\\
 =& -\frac{1}{2} \sbkt{\int_0^t \dd \tau_1 \int_0^t \dd \tau_2\bkt{ {G} \bkt{t_1-\tau_1} \nu^{(-)}\bkt{\tau_1-\tau_2} {G}\bkt{\tau_2-t_2} + {G}\bkt{t_2 - \tau_1}\nu^{(-)}\bkt{\tau_1-\tau_2} {G}\bkt{\tau_2-t_1}}}\non\\
&+\frac{i}{2} \sbkt{{G}\bkt{t_1-t_2}+{G}\bkt{t_2-t_1}}  - \aavg{ Q_h\bkt{ t_1} Q_h\bkt{t_2}}\\
= & - \nu'_{GG}\bkt{ t_1, t_2 }  + \frac{i }{2} {g}\bkt{t_1 - t_2}\mr{sign} \bkt{t_1 - t_2}
}
Substituting back in Eq.\eqref{seffxpxp}
\eqn{&\avg{\cbkt{\sint^{(+)}\sbkt{X',-\frac{1}{i}\frac{\delta}{\delta {\bf J'}}}}^2}_0 \non\\
=&  \int_0^t\dd t_1 \int_0^t\dd t_2 \sum_{p} \lambda^2k_p^2 X'(t_1)X'(t_2)\sbkt{i\eta_p^{(+)} \bkt{t_1-t_2 }\mr{sign}\bkt{t_1-t_2}-\nu_p^{(+)}\bkt{t_1-t_2}}\non\\
&\sbkt{- \nu_{GG}'\bkt{ t_1, t_2 }  - \frac{i }{2} {g}\bkt{t_1 - t_2}\mr{sign} \bkt{t_1 - t_2} }\\
\label{seffXPXP}
= & \int_0^t\dd t_1 \int_0^t\dd t_2 \sum_{p} \lambda^2k_p^2 X'(t_1)X'(t_2)\sbkt{ -i \eta_p ^{(+)} \bkt{t_1 - t_2} \nu_{GG}' \bkt{t_1 , t_2} \mr{sign} \bkt{t_1 - t_2} + \nu_p ^{(+)}\bkt{t_1 - t_2} \nu_{GG}' \bkt{t_1 , t_2}\right.\non\\ 
&\left. - \frac{1}{2} {g}\bkt{t_1 - t_2} \eta_p ^{(+)} \bkt{t_1 - t_2}- \frac{i}{2} {g}\bkt{t_1 - t_2} \nu_p ^{(+)} \bkt{t_1 - t_2}\mr{sign}\bkt{t_1-t_2}}.
}
}
\item{ \eqn{ \label{seffxxp}\avg{\sint^{(+)}\sbkt{X',-\frac{1}{i}\frac{\delta}{\delta {\bf J'}}}\sint^{(+)}\sbkt{X,\frac{1}{i}\frac{\delta}{\delta {\bf J}}}}_0 =  \int_0^t\dd t_1 \int_0^t\dd t_2 \sum_{m,n} \lambda^2\omega_m\omega_n X(t_1)X'(t_2)\,\widetilde{\zeta}_I\bkt{t_1,t_2}\,\sum_{p} \widetilde{\zeta}^{(p)}_{m,n}\bkt{t_1,t_2}
}
where
\eqn{\label{tildezetap}\widetilde{\zeta}^{(p)}_{m,n}(t_1,t_2)\equiv &\left.\frac{\delta }{\delta J_n(t_1)}\cbkt{\frac{\delta }{\delta J'_m(t_2)}\sbkt{F^{(1)}_p [J_p,J'_p]}}\right\vert_{J_p = J'_p =0}\non\\
=& -i\delta_{mp}\delta_{np}\sbkt{\eta_p^{(+)} \bkt{t_1-t_2 }\Theta\bkt{t_1-t_2}-\eta_p^{(+)} \bkt{t_2-t_1}\Theta\bkt{t_2-t_1}}\non\\
&+\delta_{mp}\delta_{np}\sbkt{\nu_p^{(+)} \bkt{t_1-t_2}\Theta\bkt{t_1-t_2}+\nu_p^{(+)}\bkt{t_2-t_1}\Theta\bkt{t_2-t_1}}\\
= & \delta_{mp}\delta_{np}\sbkt{-i\eta_p^{(+)} \bkt{t_1-t_2 } + \nu_p ^{(+)} \bkt{t_1 - t_2}},
}
and
\eqn{\label{tildezetaI}\widetilde{\zeta}_I(t_1,t_2)\equiv& \frac{\delta }{\delta J(t_1)}\cbkt{\frac{\delta }{\delta J'(t_2)}\sbkt{F^{(1)}_I [J,J']}}\non\\
 =& \frac{1}{2} \sbkt{\int_0^t \dd \tau_1 \int_0^t \dd \tau_2\bkt{ {G} \bkt{t_1-\tau_1} \nu^{(-)}\bkt{\tau_1-\tau_2} {G}\bkt{\tau_2-t_2} + {G}\bkt{t_2 - \tau_1}\nu^{(-)}\bkt{\tau_1-\tau_2} {G}\bkt{\tau_2-t_1}}}\non\\
&+\frac{i}{2} \sbkt{{G}\bkt{t_1-t_2}-{G}\bkt{t_2-t_1}}+   \aavg{ Q_h\bkt{ t_1} Q_h\bkt{t_2}}\\
= & \nu'_{GG}\bkt{t_1 ,t_2 } - \frac{i}{2}  g\bkt{t_1 - t_2} .
}
Substituting the above in Eq.\eqref{seffxxp}
\eqn{
&\avg{\sint^{(+)}\sbkt{X',-\frac{1}{i}\frac{\delta}{\delta {\bf J'}}}\sint^{(+)}\sbkt{X,\frac{1}{i}\frac{\delta}{\delta {\bf J}}}}_0 \non\\
=&  \int_0^t\dd t_1 \int_0^t\dd t_2 \sum_{p} \lambda^2k_p^2 X(t_1)X'(t_2)\sbkt{-i \eta_p ^{(+)} \bkt{t_1 - t_2} + \nu_p ^{(+)} \bkt{t_1 - t_2}}  \sbkt{\nu_{GG}'\bkt{t_1 ,t_2 } +\frac{i}{2}  g\bkt{t_1 - t_2}}\\
\label{seffXXP}
= & \int_0^t\dd t_1 \int_0^t\dd t_2 \sum_{p} \lambda^2k_p^2 X(t_1)X'(t_2)\sbkt{-i \eta_p ^{(+)} \bkt{t_1 - t_2} \nu_{GG}'\bkt{t_1 ,t_2 } + \nu_p ^{(+)} \bkt{t_1 - t_2} \nu_{GG}'\bkt{t_1 ,t_2 }\right.\non\\
  &\left.-\frac{1}{2}  g\bkt{t_1 - t_2}\eta_p ^{(+)} \bkt{t_1 - t_2}- \frac{i}{2} {g}\bkt{t_1 - t_2}\nu_p ^{(+)} \bkt{t_1 - t_2}} . 
}
}
\end{enumerate}

Putting together Eqs.\,\eqref{seffXX}, \eqref{seffXPXP}, and \eqref{seffXXP} in Eq.\eqref{seff}, we obtain the second order influence action as 
\eqn{&S_\mr{M, IF}^{(2)}\sbkt{X,X'}\non\\
= &\frac{i}{2}\sum_p\lambda^2 k_p^2\int_0^t \dd t_1\int_0^t \dd t_2\sbkt{  X(t_1) X(t_2)\cbkt{i \eta_p ^{(+)}\bkt{t_1 - t_2 } \nu_{GG}' \bkt{t_1, t_2}\mr{sign}\bkt{t_1-t_2}+ \nu_p ^{(+)} \bkt{t_1 - t_2} \nu_{GG}' \bkt{t_1, t_2} \right.\right.\non\\
&\left.\left. - \frac{1}{2} {g}\bkt{t_1 - t_2} \eta_p ^{(+)} \bkt{t_1 - t_2}+ \frac{i}{2} {g}\bkt{t_1 - t_2} \nu_p ^{(+)} \bkt{t_1 - t_2}\mr{sign}\bkt{t_1-t_2} }\right.\non\\
&\left.+  X'(t_1)X'(t_2)\cbkt{ -i \eta_p ^{(+)} \bkt{t_1 - t_2} \nu_{GG}' \bkt{t_1 , t_2} \mr{sign} \bkt{t_1 - t_2} + \nu_p ^{(+)}\bkt{t_1 - t_2} \nu_{GG}' \bkt{t_1 , t_2}\right.\right.\non\\ 
&\left. \left. - \frac{1}{2} {g}\bkt{t_1 - t_2} \eta_p ^{(+)} \bkt{t_1 - t_2}- \frac{i}{2} {g}\bkt{t_1 - t_2} \nu_p ^{(+)} \bkt{t_1 - t_2}\mr{sign}\bkt{t_1-t_2}}\right.\non\\
& \left.-2  X(t_1)X'(t_2)\cbkt{-i \eta_p ^{(+)} \bkt{t_1 - t_2} \nu_{GG}'\bkt{t_1 ,t_2 } + \nu_p ^{(+)} \bkt{t_1 - t_2} \nu_{GG}'\bkt{t_1 ,t_2 }\right.\right.\non\\
  &\left.\left.- \frac{1}{2}  g\bkt{t_1 - t_2}\eta_p ^{(+)} \bkt{t_1 - t_2}- \frac{i}{2} {g}\bkt{t_1 - t_2}\nu_p ^{(+)} \bkt{t_1 - t_2}}}}
  \eqn{  = & \frac{i}{8}\sum_p\lambda^2 k_p^2\int_0^t \dd t_1\int_0^t \dd t_2\sbkt{  \cbkt{X_\Sigma(t_1) + X_\Delta(t_1) }\cbkt{X_\Sigma(t_2) + X_\Delta(t_2) }\cbkt{i \eta_p ^{(+)}\bkt{t_1 - t_2 } \nu_{GG}' \bkt{t_1, t_2}\mr{sign}\bkt{t_1-t_2} \right.\right.\non\\
&\left.\left. + \nu_p ^{(+)} \bkt{t_1 - t_2} \nu_{GG}' \bkt{t_1, t_2}- \frac{1}{2} {g}\bkt{t_1 - t_2} \eta_p ^{(+)} \bkt{t_1 - t_2}+ \frac{i}{2} {g}\bkt{t_1 - t_2} \nu_p ^{(+)} \bkt{t_1 - t_2}\mr{sign}\bkt{t_1-t_2} }\right.\non\\
&\left.+  \cbkt{X_\Sigma(t_1) - X_\Delta(t_1) }\cbkt{X_\Sigma(t_2) - X_\Delta(t_2) }\cbkt{ -i \eta_p ^{(+)} \bkt{t_1 - t_2} \nu_{GG}' \bkt{t_1 , t_2} \mr{sign} \bkt{t_1 - t_2} \right.\right.\non\\ 
&\left. \left. + \nu_p ^{(+)}\bkt{t_1 - t_2} \nu_{GG}' \bkt{t_1 , t_2}- \frac{1}{2} {g}\bkt{t_1 - t_2} \eta_p ^{(+)} \bkt{t_1 - t_2}- \frac{i}{2} {g}\bkt{t_1 - t_2} \nu_p ^{(+)} \bkt{t_1 - t_2}\mr{sign}\bkt{t_1-t_2}}\right.\non\\
& \left.-2  \cbkt{X_\Sigma(t_1) + X_\Delta(t_1) }\cbkt{X_\Sigma(t_2) - X_\Delta(t_2) }\cbkt{-i \eta_p ^{(+)} \bkt{t_1 - t_2} \nu_{GG}'\bkt{t_1 ,t_2 } + \nu_p ^{(+)} \bkt{t_1 - t_2} \nu_{GG}'\bkt{t_1 ,t_2 }\right.\right.\non\\
  &\left.\left.- \frac{1}{2}  g\bkt{t_1 - t_2}\eta_p ^{(+)} \bkt{t_1 - t_2}- \frac{i}{2} {g}\bkt{t_1 - t_2}\nu_p ^{(+)} \bkt{t_1 - t_2}}}}
  \eqn{   = & \frac{i}{4}\sum_p\lambda^2 k_p^2\int_0^t \dd t_1\int_0^t \dd t_2\non\\
  &\sbkt{  X_\Sigma(t_1) X_\Sigma(t_2) \cbkt{ -i \eta_p ^{(+)} \bkt{t_1 - t_2} \nu_{GG}'\bkt{t_1 ,t_2 }- \frac{i}{2} {g}\bkt{t_1 - t_2}\nu_p ^{(+)} \bkt{t_1 - t_2}} \right.\non\\
  &\left.+X_\Delta(t_1) X_\Delta(t_2) \cbkt{2\nu_p ^{(+)} \bkt{t_1 - t_2} \nu_{GG}'\bkt{t_1 ,t_2 }- g\bkt{t_1 - t_2}\eta_p ^{(+)} \bkt{t_1 - t_2} -i \eta_p ^{(+)} \bkt{t_1 - t_2} \nu_{GG}'\bkt{t_1 ,t_2 }\right.\right.\non\\
  &\left.\left.-\frac{i}{2} {g}\bkt{t_1 - t_2}\nu_p ^{(+)} \bkt{t_1 - t_2}}\right.\non\\ &\left.+  X_\Sigma(t_1) X_\Delta(t_2) \cbkt{i \eta_p ^{(+)} \bkt{t_1 - t_2} \nu_{GG}'\bkt{t_1 ,t_2 }\bkt{\mr{sign}\bkt{t_1 - t_2}-1}+ \frac{i}{2} {g}\bkt{t_1 - t_2}\nu_p ^{(+)} \bkt{t_1 - t_2}\bkt{ \mr{sign}\bkt{t_1 - t_2}-1}\right.\right.\non\\ &\left.\left. + \nu_p ^{(+)} \bkt{t_1 - t_2} \nu_{GG}'\bkt{t_1 ,t_2 }- \frac{1}{2}  g\bkt{t_1 - t_2}\eta_p ^{(+)} \bkt{t_1 - t_2} }\right.\non\\
&  \left.+  X_\Delta(t_1) X_\Sigma(t_2)\cbkt{i \eta_p ^{(+)} \bkt{t_1 - t_2} \nu_{GG}'\bkt{t_1 ,t_2 }\bkt{\mr{sign}\bkt{t_1 - t_2}+1}+ \frac{i}{2} {g}\bkt{t_1 - t_2}\nu_p ^{(+)} \bkt{t_1 - t_2}\bkt{ \mr{sign}\bkt{t_1 - t_2}+1}\right.\right.\non\\ &\left.\left. - \nu_p ^{(+)} \bkt{t_1 - t_2} \nu_{GG}'\bkt{t_1 ,t_2 }-\frac{1}{2}  g\bkt{t_1 - t_2}\eta_p ^{(+)} \bkt{t_1 - t_2} }}\\
  = & \frac{i}{4}\sum_p\lambda^2 k_p^2\int_0^t \dd t_1\int_0^t \dd t_2\sbkt{X_\Delta(t_1) X_\Delta(t_2) \cbkt{ 2\nu_p ^{(+)} \bkt{t_1 - t_2} \nu_{GG}'\bkt{t_1 ,t_2 } - g\bkt{t_1 - t_2} \eta_p ^{(+)} \bkt{t_1 - t_2}}\right.\non\\
&   \left.+  X_\Delta(t_1) X_\Sigma(t_2)\cbkt{2i\eta_p ^{(+)} \bkt{t_1 - t_2} \nu_{GG}'\bkt{t_1 ,t_2 }+ i {g}\bkt{t_1 - t_2}\nu_p ^{(+)} \bkt{t_1 - t_2} }\Theta\bkt{t_1 - t_2}},
}
where we have defined $X_\Sigma\bkt{t} \equiv X\bkt{t}+ X'\bkt{t}$, and $X_\Delta\bkt{t} \equiv X\bkt{t}- X'\bkt{t}$, and for the last line we have used the property $\int_{0}^{t}dt_{1}\int_{0}^{t}dt_{2}f\bkt{t_{1}}f\bkt{t_{2}}F\bkt{t_1 - t_2}H\bkt{t_1,t_2}=0$ provided that $F\bkt{t_2 - t_1}=-F\bkt{t_1 - t_2}$ and $H\bkt{t_2,t_1}=H\bkt{t_1,t_2}$. Using the definitions of the dissipation and noise kernels for the (+) bath as in Eq.\eqref{ETA+} and Eq. \eqref{NU+}
\eqn{S_\mr{M, IF}^{(2)}=  \int_0^t \dd t_1\int_0^t \dd t_2&\sbkt{\frac{i}{4} X_\Delta(t_1) X_\Delta(t_2) \cbkt{ 2\nu ^{(+)} \bkt{t_1 - t_2} \nu_{GG}'\bkt{t_1 ,t_2}- g\bkt{t_1 - t_2} \eta_p ^{(+)} \bkt{t_1 - t_2} }\right.\non\\
&  \left.- \frac{1}{4} X_\Delta(t_1) X_\Sigma(t_2)\cbkt{ {g}\bkt{t_1 - t_2}\nu ^{(+)} \bkt{t_1 - t_2} +2\eta ^{(+)} \bkt{t_1 - t_2} \nu_{GG}'\bkt{t_1 ,t_2 }}\Theta\bkt{t_1 - t_2}},}
which reduces to Eq.\,\eqref{EffAct2nd}.

\section{Fourier 
transform of the dissipation and noise kernels}
\label{App:kernels}

The Fourier 
transform of the dissipation and noise kernels corresponding to the baths associated with the scalar field $\{\eta^{(\pm)},\nu^{(\pm)}\}$ are
\eqn{ \label{etapmw}
\bar\eta^{(\pm)}(\omega)=&i{\rm Im}\sbkt{\bar \eta^{(\pm)}(\omega)}=-i\frac{\pi}{2}\sbkt{ J^{(\pm)}(\omega)- J^{(\pm)}(-\omega)},\\
\label{nupmw}
\bar\nu^{(\pm)}(\omega)=&\pi\coth\left(\frac{\omega}{2k_{B}T_{F}}\right)\sbkt{ J^{(\pm)}(\omega)-   J^{(\pm)}(-\omega)}
.
}
%
From these expressions, we immediately obtain a fluctuation-dissipation relation for the $ \bkt{\pm}$ baths as in Eq.\eqref{FDRBathPM}.

We now consider the  kernel $\nu_{GG}(t_1,t_2)$ in the late-time limit (Eq.\eqref{NuGGLongTime}) as follows:


%
\eqn{\nu_{GG}(t_1,t_2)=\int_{-\infty}^{+\infty}\frac{d\omega}{2\pi}\bar{\nu}^{(-)}\bkt{\omega}\int_{0}^{t_{1}}\dd\tau_{1}~G
\bkt{t_{1}-\tau_{1}}e^{-i\omega\tau_{1}}\int_{0}^{t_{2}}\dd\tau_2~G
\bkt{t_{2}-\tau_{2}}e^{i\omega\tau_{2}},
\label{nuGGFourierFiniteTime}
}
where we have written $\nu^{(-)}(t)$ in terms of its Fourier transform. The time integrals on a finite interval are convolutions that can be re-cast in terms of the Laplace transform of $G$ (defined as $\tilde F(z)=\int_{0}^{+\infty}{dt} e^{-z t} F(t)$) via its inverse as:
\eqn{\int_{0}^{t}\dd\tau~G
\bkt{t-\tau}e^{\pm i\omega\tau}=\int_{l-i\infty}^{l+i\infty}\frac{dz}{2\pi i}~e^{zt}\frac{\tilde{G
}(z)}{(z\mp i\omega)},
}
where $l$ has to be larger than the real parts of each of the poles of $\tilde{G
}$ to ensure causality. Specifically, as $\tilde{G
}$ has poles $\{z_{k}\}$ with negative or vanishing real parts (${\rm Re}(z_{k})<0$), implying $l>0$. Using Cauchy's theorem, the causal property of the retarded propagator allow us to express the convolution in terms of a sum of the residue of the poles of $\tilde{G}(z)/(z\mp i\omega)$   as:
\eqn{\int_{0}^{t}\dd\tau~G
\bkt{t-\tau}e^{\pm i\omega\tau}=e^{\pm i\omega t}\bar{G}(\mp\omega)
+\sum_{k}e^{z_{k} t}\frac{\mathcal{R}_{k}}{(z_{k}\mp i\omega)},
}
with $\mathcal{R}_{k}\equiv{\rm Res}[\tilde{G
}(z),z_{k}]$ and where we have used the relation between the Fourier and Laplace transforms of causal functions $\tilde{G}(\pm i\omega)=\bar{G}(\mp \omega)$.

Considering that ${\rm Re}(z_{k})<0$, it is easy to realize that in the large-time limit $t\rightarrow+\infty$:
\eqn{\int_{0}^{t}\dd\tau~G
\bkt{t-\tau}e^{\pm i\omega\tau}\xrightarrow{t\rightarrow+\infty}e^{\pm i\omega t}\bar{G}(\mp\omega)
,
}
which immediately allow us to obtain for the long-time limit of Eq.\eqref{nuGGFourierFiniteTime}:
\eqn{\nu_{GG}(t_1,t_2)\rightarrow\int_{-\infty}^{+\infty}\frac{d\omega}{2\pi}e^{-i\omega(t_{1}-t_{2})}\left|\bar{G}(\omega)
\right|^{2}\bar{\nu}^{(-)}\bkt{\omega},
\label{nuGGFourierLongTime}
}
where we have used that $\bar{G}\bkt{-\omega}=\bar{G}^{*}\bkt{\omega}$ 
given that $G  $ is real. Similarly, we note that the last expression is a function of $t_{1}-t_{2}$ only, which is generally not the case during the course of the evolution. In the late-time limit, we can define the Fourier transform of the kernel as $\bar \nu_{GG}(\omega)\equiv|\bar{G}\bkt{\omega}
|^{2}\bar{\nu}^{(-)}\bkt{\omega}$. We note that $\bar\nu_{GG}(\omega)$ is real and even, as expected.

Furthermore, considering Eq.\eqref{Gret}, we have that the Fourier 
transform of the retarded propagator reads:
\eqn{\bar {G}(\omega)
=\frac{1}{m\omega_{0}^{2}-m\omega^{2}+2\bar {\eta}^{(-)}(\omega)
}.
}

Noticing that $\bar{\eta}^{(-)}(\omega)
$ has real and imaginary parts, we leave the real part as a renormalization of the frequency of the center of mass, while the imaginary part accounts for the dissipation. 
After implementing this, we consider the renormalized frequency $\omega_{R}^{2}\equiv\omega_{0}^{2}+2{\rm Re}[\bar{\eta}^{(-)}(\omega)]/m$, so we can re-cast the Fourier 
transform of the retarded propagator as:
\eqn{\bar{G}(\omega)
=\frac{1}{m\omega_{R}^{2}-m\omega^{2}+2i{\rm Im}[\bar{\eta}^{(-)}(\omega)
]},
}
from which we obtain that ${\rm Im}[\bar {G}(\omega)]=-2|\bar{G}(\omega)|^{2}{\rm Im}[\bar{\eta}^{(-)}(\omega)]$. Combining this with the fluctuation-dissipation relation for the bath $(-)$ of Eq.\eqref{FDRBathPM}, we can directly arrive at 
the fluctuation-dissipation relation between $\bar\nu_{GG}\bkt{\omega}$ and $\bar{G}$ as in Eq.\,\eqref{FDRnuGG}.

\end{document}